\documentclass[twocolumn]{emulateapj}
\usepackage{amsmath}
\usepackage{nicefrac}
\usepackage[none]{hyphenat}
\def\Lsun{L$_{\odot}$}
\def\Rsun{R$_{\odot}$}
\def\Msun{M$_{\odot}$}

\begin{document}

\title{Dynamical Formation of Close Binaries during the Pre-main-sequence Phase}

\author{Maxwell Moe\altaffilmark{1,2} \& Kaitlin M.  Kratter\altaffilmark{1}}

\altaffiltext{1}{Steward Observatory, University of Arizona, 933~N.~Cherry~Ave.,~Tucson,~AZ 85721,~USA}

\altaffiltext{2}{Einstein Fellow}

\begin{abstract}
 Solar-type binaries with short orbital periods ($P_{\rm close}$~$\equiv$~1\,-\,10~days; $a$~$\lesssim$~0.1~AU) cannot form directly via fragmentation of molecular clouds or protostellar disks, yet their component masses are highly correlated, suggesting interaction during the pre-main-sequence (pre-MS) phase.  Moreover, the close binary fraction of pre-MS stars is consistent with that of their MS counterparts in the field ($F_{\rm close}$~$=$~2.1\%).  Thus we can infer that some migration mechanism operates during the early pre-MS phase ($\tau$~$\lesssim$~5~Myr) that reshapes the primordial separation distribution. We test the feasibility of this hypothesis by carrying out a population synthesis calculation which accounts for two formation channels: Kozai-Lidov (KL) oscillations and dynamical instability in triple systems.  Our models incorporate (1) more realistic initial conditions compared to previous studies, (2) octupole-level effects in the secular evolution, (3) tidal energy dissipation via weak-friction equilibrium tides at small eccentricities and via non-radial dynamical oscillations at large eccentricities, and (4) the larger tidal radius of a pre-MS primary.  Given a 15\% triple star fraction, we simulate a close binary fraction from KL oscillations alone of $F_{\rm close}$~$\approx$~0.4\% after $\tau$~=~5~Myr, which increases to $F_{\rm close}$~$\approx$~0.8\% by $\tau$~=~5~Gyr. Dynamical ejections and disruptions of unstable coplanar triples in the disk produce solitary binaries with slightly longer periods $P$~$\approx$~10\,-\,100~days. The remaining $\approx$60\% of close binaries with outer tertiaries, particularly those in compact coplanar configurations with log\,$P_{\rm out}$\,(days)~$\approx$~2\,-\,5 ($a_{\rm out}$~$<$~50~AU), can be explained only with substantial extra energy dissipation due to interactions with primordial gas. 
\end{abstract}


\section{Introduction}

Close binaries can evolve to produce a variety of astrophysical phenomena, including Type~Ia supernovae, novae, blue stragglers, Type~Ib/c supernovae, short gamma-ray bursts, mergers of compact objects, and sources of gravitational waves (\citealt{DeMarco2017}, and  references therein).  Despite their importance, the dominant formation mechanism of close binaries remains unknown.  It is generally believed that close stellar companions cannot form {\it in situ} but instead initially fragment on large, core scales of several 100s to 1,000s of AU or within the primordial circumstellar disk at separations of several 10s of AU (see reviews by \citealt{Tohline2002} and \citealt{Kratter2011}).  The initial hydrostatic stellar core first forms with a radius of $\sim$\,5~AU \citep{Larson1969}, and fragmentation of the hydrostatic core while it is contracting is extremely unlikely \citep{Bate1998,Bate2011}.  Binaries are therefore not expected to form directly with separations smaller than $\sim$\,10~AU \citep{Boss1986,Bate1998,Bate2009}.  Some process for orbital evolution is then required to bring the binary to shorter periods.  Likely candidates include  migration through the circumbinary disk due to hydrodynamical forces, orbital decay from protostellar accretion, gravitational interactions in a dynamically unstable triple-star system, or secular evolution in a triple star, such as Kozai-Lidov cycles, coupled with tidal friction \citep{Kozai1962,Lidov1962,Artymowicz1983,Artymowicz1991,Bate1995,Bate1997,Kiseleva1998,Bate1998,Reipurth2001,Bate2002,Bate2009}.

\citet{Tokovinin2006} showed that (80\,$\pm$\,6)\% of solar-type main-sequence (MS) binaries with periods $P$~$<$~7~days have outer tertiaries with $q$ = $M_3$/$M_1$ $>$ 0.2.  Meanwhile, they found that only $\approx$30\% of solar-type binaries with slightly longer periods $P$~$>$~20~days have such tertiary components. After correcting for selection effects, \citet{Moe2017} recently demonstrated the very close binary fraction is directly proportional to the overall triple/quadruple star fraction, independent of primary mass.  They also found that $\approx$15\% of triples have inner binary periods $P_{\rm in}$~$<$~7~days.  These observations demonstrate that triple stars play a key role in the formation of very close binaries.

However, these particular observations do not tell us precisely {\it how} triples lead to the formation of close binaries. \citet{Bate2002} and \citet{Bate2009} suggest the rapid dynamical unfolding of an initially unstable triple coupled with significant energy dissipation within the primordial disk can produce a close binary during the pre-MS phase.  While their hydrodynamical simulations can reproduce certain observed features of the binary population, their models cannot yet produce very close binaries with $a_{\rm in}$~$<$~0.1~AU (see Fig.~18 in \citealt{Bate2009} and \citealt{Bate2012}).  They argue the finite resolution of their hydrodynamical simulations limits their ability to form very close binaries.  Future modeling is required to determine if finer resolution can indeed reproduce the necessary number of close binaries via this triple-star~+~disk rapid hydrodynamical formation channel. Given that the typical fragmentation scale in disks is of order $\sim$\,50~AU, a substantial amount of orbital energy and angular momentum must be removed to reach separations of $\lesssim$\,0.1~AU.

An alternative explanation for close binaries with outer tertiaries involves secular evolution of triple stars via Kozai-Lidov (hereafter KL) oscillations \citep{Kiseleva1998}.  In this scenario, the outer tertiary gradually pumps the eccentricity of the inner binary to large values after many orbital timescales, at which point tidal interactions cause the inner binary to decay to shorter periods.  KL oscillations may also explain the dynamical formation of hot Jupiters, whereby secular interactions between a Jupiter-mass planet and a wide binary companion cause the planet to migrate inward \citep{Wu2003,Fabrycky2007,Naoz2012}.  KL driven evolution also explains secondaries that have orbits inclined with respect to the rotation axes of their primary stars \citep{Triaud2010,Albrecht2009,Albrecht2012,Anderson2017}.  

To investigate the efficiency of forming close binaries via KL cycles, \citet{Fabrycky2007} utilized a Monte Carlo population synthesis technique.  They drew both inner and outer companions from the observed companion distribution of solar-type MS binaries \citep{Duquennoy1991}, checked for dynamical stability of each triple, and then evolved each system for 10 Gyr according to the quadrupole-level approximation for KL oscillations.  More recently, \citet{Naoz2014} performed a similar calculation, using the same initial conditions and simulating for 10 Gyr, but included the effects of octupole-level secular evolution.  Both of these studies found a measurable enhancement in the close binary population, demonstrating that KL cycles coupled with tidal friction can in principle produce close binaries within the MS lifetime of solar-type stars.

However, the observed binary period distribution of solar-type MS systems in the field ($\tau$~$\approx$~5~Gyr) is already significantly dynamically processed \citep{Kroupa1995,Goodwin2005,Marks2011}. It is therefore inconsistent to select initial inner and outer periods of triples from the currently observed solar-type MS binary period distribution.  In particular, a significant fraction of the close inner binaries with $P_{\rm in}$~$<$~10~days simulated in \citet{Fabrycky2007} and \citet{Naoz2014} derived from only slightly longer initial orbital periods $P_{\rm in,0}$~$=$~10\,-\,100~days ($a_{\rm in,0}$~$=$~0.1\,-\,0.5~AU).  Both observations of pre-MS multiple star systems and theoretical simulations indicate that stellar companions initially form at scales of several 10s of AU via disk fragmentation and several 100s to 1,000s of AU via core fragmentation  \citep{Boss1986, Bonnell1994, Bate1995, Bate1998, Tohline2002, Matzner2005, Kratter2008, Bate2009, Offner2010, Kratter2011, Pineda2015, Tobin2016a,Tobin2016b}. The initial conditions of triple stars should be consistent with these two modes of forming stellar companions.  

Moreover, triple-star secular evolution that requires up to 10~Gyr of the MS lifetime of solar-type systems cannot reproduce two important features of the observed close binary star population.  First, short-period binaries exhibit a measurable excess fraction of twin components with mass ratios close to unity \citep{Tokovinin2000,Halbwachs2003}.  After accounting for selection biases, \citet{Moe2017} estimate that $\approx$(20\,-\,30)\% of solar-type binaries with $P$~$<$~20~days have mass ratios $q$~=~$M_2$/$M_1$~=~0.9\,-\,1.0.  This sharp excess twin fraction demonstrates that close binaries coevolved in a manner that drove the mass ratio toward unity, either through stable mass transfer during the early pre-MS phase or shared accretion in the primordial circumbinary disk  \citep{Kroupa1995,Bate1997,Tokovinin2000,Bate2000,Bate2012}.  Even close massive binaries with $M_1$~$>$~10\,\Msun\ exhibit an excess twin fraction $F_{\rm twin}$~$\approx$~10\,-\,20\% \citep{Pinsonneault2006,Moe2017}, suggesting at least some companions to massive stars migrate inward during the early pre-MS phase (see also \citealt{Moe2015b}). While KL oscillations coupled with tidal evolution may preferentially lead to closer binaries with larger mass ratios, these processes alone cannot reproduce the observed steep, nearly discontinuous excess fraction of twin components.  

Second, the close binary fraction of solar-type stars appears to be universal, i.e., relatively independent of age or environment (see reviews by \citealt{Duchene2013} and \citealt{Moe2017}).  In particular, the close binary fraction of solar-type MS stars in the field \citep{Duquennoy1991,Raghavan2010}, of solar-type MS stars in various open clusters spanning a wide range of ages and densities \citep{Geller2012,Leiner2015}, and of young $\tau$~$\approx$~1\,-\,5~Myr pre-MS T~Tauri stars \citep{Mathieu1994,Melo2003} are all consistent with each other (see Fig.~41 in \citealt{Moe2017}).  By taking an average of these observations, we find that $F_{\rm close}$~=~2.1\% of solar-type MS stars have companions with $P_{\rm close}$~$\equiv$~1\,-\,10~days ($a$~$\lesssim$~0.1~AU). If close binaries originally formed at wider separations, then observations dictate the dynamical evolution toward shorter periods must have predominantly occurred during the early pre-MS phase when there was still a disk ($\tau$~$\lesssim$~$\tau_{\rm disk}$~$\approx$~5~Myr).  Such early migration through and coevolution in the primordial disk could also explain the observed excess twin fraction. 

The majority of close binaries must have evolved to their currently observed configurations during the pre-MS phase, which is three orders of magnitude shorter than the 10 Gyr simulations conducted by \citet{Fabrycky2007} and \citet{Naoz2014}.  The larger radius of the pre-MS primary may potentially make the formation of close binaries via KL oscillations and tidal interactions more efficient at early times.  The purpose of this study is to perform Monte Carlo simulations of triple star dynamical evolution by implementing more realistic initial conditions and different prescriptions for tidal energy dissipation.  We can then more reliably determine whether a combination of dynamical instability and KL oscillations coupled with tidal interactions during the short-lived pre-MS phase can reproduce the observed close binary population. In \S2, we simulate the evolution of triples born in dynamically stable configurations in which the inner  binaries form via disk fragmentation at intermediate separations and the outer tertiaries derive from core fragmentation on large spatial scales.  Some of these triples form close binaries via KL cycles and tidal friction, and we discuss the results of our simulations in \S3.  In \S4, we consider the evolution of triples born in dynamically unstable configurations, focusing on systems where both inner and outer companions fragment within the disk.  In \S5, we summarize our results and discuss the implications for the formation of close binaries and hot Jupiters. 

\section{Dynamically Stable Triples \\ from Disk + Core Fragmentation}

\subsection{Initial Conditions}

\subsubsection{Companion Distributions}

Using a Monte Carlo technique, we generate the initial properties of triple stars in a manner that is consistent with both the theoretical models of multiple star formation and the observed multiplicity statistics of pre-MS systems as cited and discussed in \S1.  For our simulations in this section (Models A1\,-\,A11, see Table 1), we assume the inner binary forms via disk fragmentation at intermediate separations,  the outer tertiary forms via core fragmentation at wide separations, and the resulting triple-star system is dynamically stable.  For our baseline model (Model A1), we select the initial logarithmic separation log $a_{\rm in,0}$ (AU) of the inner binary from a Gaussian distribution with mean $\mu_{\rm log\,a;\,in}$~=~1.5 (30~AU) and dispersion of $\sigma_{\rm log\,a;\,in}$~=~0.8.  The peak at $a_{\rm in,0}$~=~30~AU in the inner binary period distribution matches the peak in the overall companion distribution of solar-type MS binaries \citep{Duquennoy1991, Raghavan2010}.  We then select the outer tertiary from a Gaussian distribution with mean $\mu_{\rm log\,a;\,out}$~=~2.8 (600~AU) and dispersion of $\sigma_{\rm log\,a;\,out}$~=~1.0. We generate companions solely across the interval 0.5~AU~$<$~$a_{\rm in,0}$ $<$ $a_{\rm out,0}$~$<$~30,000~AU, and we check for dynamical stability of each triple once we select the other physical parameters of the system (see below).  To assess the robustness of our model, we consider initial separations of the inner binaries to be systematically closer $\mu_{\rm log\,a;\,in}$~=~1.0 (10~AU; Model~A2) and wider $\mu_{\rm log\,a;\,in}$~=~2.0  (100~AU; Model~A3).  Similarly, we vary the outer tertiary separation distribution to be systematically smaller $\mu_{\rm log\,a;\,out}$~=~2.3  (200~AU; Model~A4) and larger $\mu_{\rm log\,a;\,out}$~=~3.3 (2,000~AU; Model~A5).   Our baseline model A1 is not necessarily the most correct model but nevertheless serves as an adequate reference for comparing the changes in the simulated close binary populations as a function of different initial conditions and prescriptions for secular evolution (see \S3).

We assume the physical processes of disk and core fragmentation are relatively independent \citep{Kraus2011}.  We therefore expect the angular momentum vectors of the corresponding inner and outer orbits to be randomly oriented.  In fact, there are several known cases of pre-MS star systems in which the wide companions with $a$~$\gtrsim$~300~AU are markedly misaligned with the primordial disks of the primaries \citep{Koresko1998,Stapelfeldt1998,Kang2008,Jensen2014,Williams2014,Brinch2016,Fernandez2017,Lee2017}.  Moreover, triples with very wide tertiaries $a_{\rm out}$~$\gtrsim$~1,000~AU are generally observed to have random orientations \citep[][references therein]{Sterzik2002,Tokovinin2017}. To generate random orientations, we select the total mutual inclination $i_{\rm tot,0}$ between the orbits of the inner binary and outer tertiary from the probability distribution $p$~$\propto$~sin\,$i_{\rm tot,0}$ across the interval $i_{\rm tot,0}$ = 0$^{\circ}$\,-\,180$^{\circ}$.

We set the primary mass to be $M_1$ = 1\,\Msun\ for all the triples in our simulations.  The mass-ratio distribution of solar-type MS binaries with intermediate orbital periods is broadly consistent with a uniform distribution \citep{Raghavan2010, Moe2017}.  We select the mass ratio $q_2$ = $M_2$/$M_1$ of the inner binary from a uniform distribution across the interval $q_2$ = 0.1\,-\,1.0.  Meanwhile, the mass-ratio distribution of wide companions, including outer tertiaries, are weighted toward smaller mass ratios \citep{Lepine2007, Moe2017}.  We select the mass ratio $q_3$~=~$M_3$/$M_1$ from a probability distribution $p$~$\propto$~$q_3^{-1}$ across the interval $q_3$~=~0.1\,-\,1.0, which adequately describes the observations.  

For our inner companions that form via disk fragmentation, the initial eccentricity distribution is uncertain.   In our baseline model, we assume the inner binaries initially have nearly circular orbits, i.e., $e_{\rm in,0}$~=~0.01. Orbital circularization could plausibly occur due to interactions with the massive, natal disks, though orbits will initially be slightly more eccentric \citep{Stamatellos2009,Kratter2010}.  We thus also model an inner binary eccentricity distribution according to $p$~$\propto$~$e_{\rm in,0}^{-0.8}$ across the interval $e_{\rm in,0}$~=~0.01\,-\,0.99 (Model A6).  For the outer tertiaries that form via core fragmentation, the initial eccentricities can be much larger.  In our baseline model, we select the initial eccentricity of the outer orbit from a uniform distribution across the interval $e_{\rm out,0}$~=~0.01\,-\,0.99.  We also consider a thermal eccentricity distribution $p$ $\propto$ $e_{\rm out,0}$ \citep{Ambartsumian1937, Heggie1975, Kroupa2008} across $e_{\rm out}$~=~0.01\,-\,0.99 (Model~A7). Finally, we assume random orbital configurations, and so select initial arguments of periastron $\omega_{\rm in,0}$ and $\omega_{\rm out,0}$ independently from uniform distributions across 0$^{\circ}$\,-\,360$^{\circ}$.

\subsubsection{Stable vs. Unstable Triples}

Once the physical parameters of each triple are determined, we check for dynamical stability according to the criterion \citep{Mardling2001}:

\footnotesize
\begin{align}
 \Big( \frac{a_{\rm out,0}}{a_{\rm in,0}} \Big)_{\rm crit} =  & \frac{2.8}{1 - e_{\rm out,0}} \Big( 1 - 0.3\frac{i_{\rm tot,0}}{180^{\circ}} \Big) \nonumber \\
& \Big[ \frac{(1+q_{\rm out}) (1 + e_{\rm out,0})}{\sqrt{1 - e_{\rm out,0}}} \Big]^{\nicefrac{2}{5}},
\end{align}
\normalsize

\noindent where $q_{\rm out}$ = $M_3$/($M_1$ + $M_2$).  If a simulated triple is dynamically unstable, we generate a new system.  In our baseline model, $\approx$30\% of the simulated triples have to be regenerated in order to satisfy the dynamical stability criterion.   The dynamically unstable triples cluster near $a_{\rm in,0}$~$\sim$~$a_{\rm out,0}$~$\sim$~100~AU at the intersection of the tails of our inner and outer initial separation distributions.  After removing the unstable systems, the true distributions of $a_{\rm in,0}$ and $a_{\rm out,0}$ of our dynamically stable triples are slightly different than the generating functions used to produce the population (see Fig.~1).

Excluding the above population of initially unstable triples with $a_{\rm in,0}$~$\sim$~$a_{\rm out,0}$~$\sim$~100~AU has little impact on our results. If dynamical evolution always led to the ejection of one component from the system,  the resulting population of solitary binaries would follow a thermal eccentricity distribution $f_{\rm e}$\,d$e$~=~2$e$\,d$e$ and have a separation distribution that peaks at $a$~$\approx$~0.4$a_{\rm in,0}$~$\approx$~40~AU (\citealt{Valtonen2006}; see also \S4).  The solitary binaries would be too wide with not enough systems at sufficiently large eccentricities to evolve toward shorter separations via tides.  In this dynamical ejection scenario, a negligible fraction of initially unstable triples would produce close binaries, and so close binaries would derive almost exclusively from triples born in stable configurations. Considering 30\% of the triples initially generated in our baseline model are in unstable configurations, then the overall formation rate of close binaries in this scenario would be 70\% the rate in our baseline model. 

On the other extreme, if the initially unstable triples all evolve into marginally stable configurations with $a_{\rm out}$/$a_{\rm in}$~$\approx$~3\,-\,10, they would fall into the regime where octupole-level effects in the subsequent KL oscillations are most prominent (see \S2.2).  We show in \S3 that triples with $a_{\rm out}$/$a_{\rm in}$~$\approx$~3\,-\,10 are about twice as likely to produce close binaries via KL cycles compared to triples that are in largely hierarchical configurations $a_{\rm out}$/$a_{\rm in}$~$\approx$~10\,-\,100.  Yet in this scenario, the close binary formation rate would be 0.7\,$+$\,0.3$\times$2~=~1.3 times the rate in our baseline model.  

Between these two extremes, many of the unstable triples will evolve into largely hierarchical configurations $a_{\rm out}$/$a_{\rm in}$~$\approx$~10\,-\,100 and produce close binaries at a rate comparable to our baseline model.  In the end, the net result is that a population of 100\% dynamically stable triples (as in our main simulations) and a population of 70\% stable / 30\% unstable triples (as in our baseline model if we were not to regenerate unstable systems) will produce close binaries at approximately the same rate.  For our Monte Carlo simulations, it is therefore sufficient to regenerate triples until all the systems are dynamically stable.

\begin{figure}[t!]
\centerline{
\includegraphics[trim=0.7cm 0.3cm 0.8cm 0.2cm, clip=true, width=3.7in]{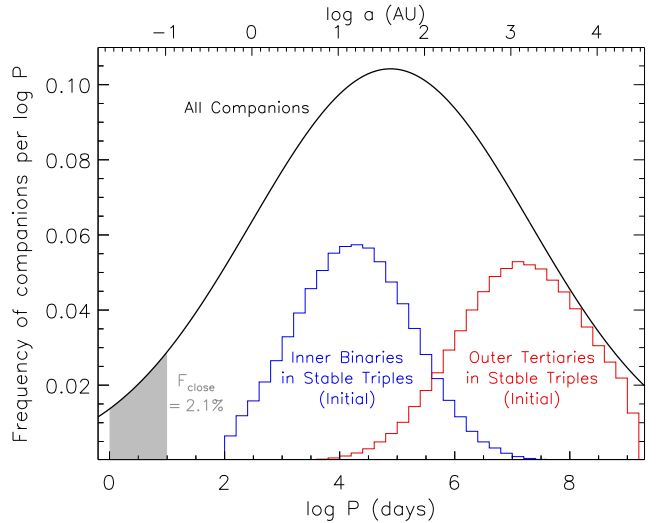}}
\caption{Frequency of companions per decade of orbital period~$P$.  We show the initial period distributions of inner binaries (blue) and outer tertiaries (red) in dynamically stable triples in our baseline model normalized to a $F_{\rm triple}$~=~15\% triple star fraction.  We also display the log-normal period distribution of companions to solar-type MS primaries (black), which provides a close binary fraction of $F_{\rm close}$ = 2.1\% across $P$~$=$~1.0\,-\,10~days (grey).  At these short periods, the close binary fraction of pre-MS T~Tauri stars ($\tau$~$\approx$~1\,-\,5~Myr) and solar-type MS stars in the field ($\tau$~$\approx$~5~Gyr) are consistent with each other.}
\end{figure}

\subsubsection{Normalization and Monte Carlo Initialization}

The fraction of solar-type field MS stars that are members of triples or higher-ordered multiples is $F_{\rm triple}$~$\approx$~(11\,-\,13)\%, depending on the survey and sensitivity toward small mass ratios \citep{Raghavan2010, Tokovinin2014, Moe2017}.  However, the frequency of wide companions to solar-type pre-MS stars is $\approx$\,2\,-\,3 times larger than the frequency of wide companions to solar-type field MS stars \citep{Ghez1993, Duchene2007, Connelley2008, Tobin2016a}.  A significant fraction of these wide pre-MS companions are actually outer tertiaries in triples \citep{Duchene2007,Connelley2008,Moe2017}, and so the pre-MS triple-star fraction $F_{\rm triple}$~$\approx$\,20\% is definitively larger.  Many of the wide companions to pre-MS stars are loosely gravitationally bound and will be ejected by the zero-age MS due to dynamical interactions with surrounding nearby stars \citep{Kroupa1995,Goodwin2005,Marks2011,Moe2017}.   Nevertheless, wide tertiary companions may induce several KL oscillations before they are dynamically disrupted.  We therefore assume that $F_{\rm triple}$~=~15\% of solar-type pre-MS primaries are initially born in dynamically stable triples with $q_2$~$>$~0.1 and $q_3$~$>$~0.1, which is only slightly larger than that currently observed in the field.  

We generate 2$\times$10$^4$ dynamically stable triples for each of our simulations. In Fig.~1, we display the initial inner and outer period distributions of dynamically stable triples in our baseline model. We also show in Fig.~1 the solar-type MS companion period distribution, which is well described by a log-normal distribution with mean $\mu_{\rm logP}$ (days) = 4.9, dispersion $\sigma_{\rm logP}$ (days) = 2.4, and overall companion frequency of 0.6 companions per primary across $-$0.3 $<$ log\,$P$\,(days) $<$ 9.3 \citep{Duquennoy1991,Raghavan2010,Moe2017}.  This solar-type MS period distribution reproduces the adopted close binary fraction $F_{\rm close}$ = 0.021 across $P$~=~1.0\,-\,10~days.  The long-period tail of outer tertiaries nearly coincides with the long-period tail of the solar-type MS companion distribution, both in terms of functional form and normalization.  This is consistent with observations, which show that nearly all very wide companions are actually tertiaries in hierarchical triples \citep{Raghavan2010,Law2010,Moe2017}.   Most importantly, we do not initiate our models with any triples that have inner binary separations $a_{\rm in,0}$~$<$~0.5~AU ($P_{\rm in,0}$~$<$~100~days), in clear contrast with the previous simulations conducted by \citet{Fabrycky2007} and \citet{Naoz2014}.

We adopt the solar-metallicity pre-MS and MS stellar evolutionary tracks from the Dartmouth Stellar Evolution Database \citep{Dotter2008}.  Our $M_1$~=~1\,\Msun\ primary starts with $R_1$~$\approx$~10\,\Rsun\ at $\tau$~=~10$^3$~yr and then contracts to $R_1$~$\approx$~2\,\Rsun\ by $\tau$~=~1~Myr.  The star then reaches its zero-age MS radius of $R_1$~=~0.9\,\Rsun\ at $\tau$~$\approx$~30~Myr. 

The initial conditions of pre-MS stellar evolutionary tracks are highly uncertain, and therefore the radial evolution of pre-MS stars at very young ages $\tau$~$\lesssim$~0.3~Myr is significantly model dependent (e.g., compare non-accreting pre-MS solar-metallicity tracks from \citealt{Baraffe1998,Siess2000,Dotter2008,Tognelli2011}).  Moreover, the cited studies model the evolution of pre-MS stars with constant mass, but \citet{Hosokawa2009} find that 1\,\Msun\ solar-metallicity stars actually accrete most of their mass while relatively small, i.e., $R$~$\lesssim$~2\,\Rsun.  Only a handful of $\approx$\,1\,\Msun\ pre-MS stars in eclipsing binaries with moderate radii $R$~$\approx$~2\,-\,3\,\Msun\ have been measured \citep{Torres2010,Moe2015b}.  Larger solar-mass pre-MS stars with $R$~$\gtrsim$~3\,\Rsun\ have yet to be identified, but it is possible they are deeply embedded in their birth clouds and therefore difficult to detect. To investigate the impact of the radial evolution of pre-MS stars on the formation of close binaries, we consider a simulation (Model A8) in which the pre-MS star has a maximum radius of $R_1$~=~3~\Rsun.  In this model, the pre-MS primary has a constant radius $R_1$~=~3~\Rsun\ for the first $\tau$~$\approx$~0.3~Myr, and then subsequently contracts according to the Dartmouth \citep{Dotter2008} pre-MS tracks.

\subsection{Kozai-Lidov Oscillations}

\subsubsection{Quadrupole-level Approximations}

If the initial mutual inclination between the inner and outer orbits satisfies a well-defined criterion (see below), then the inner binary will cycle between small and large eccentricities on the KL timescale \citep{Kiseleva1998,Antognini2015}:

\footnotesize
\begin{equation}
 \tau_{\rm KL} = \frac{8}{15 \pi} \frac{M_1 + M_2 + M_3}{M_3} \frac{P_{\rm out}^2}{P_{\rm in}} (1 - e_{\rm out}^2)^{\nicefrac{3}{2}}.
\end{equation}
\normalsize

\noindent In our baseline model, 13\% of triples have $\tau_{\rm KL}$~$<$~0.1\,Myr, and 30\% have $\tau_{\rm KL}$~$<$~1\,Myr.  

\citet{Naoz2016} recently reviewed the secular equations that describe KL oscillations.  She illustrated octupole-level effects such as flipping the orbit of the inner binary from prograde ($i_{\rm tot}$~$<$~90$^{\circ}$) to retrograde ($i_{\rm tot}$~$>$~90$^{\circ}$) with respect to the outer orbit \citep[see also][]{Li2014}.  For our stellar triples with components $M_2$~$>$~0.1$M_1$ and $M_3$~$>$~0.1$M_1$ of comparable mass, however, octupole-level effects are negligible over a broad parameter space of orbital configurations \citep{Teyssandier2013,Liu2015,Naoz2016,Anderson2017}.  Most importantly, octupole-level effects occur after several and sometimes hundreds of KL timescales.  We are mainly interested in secular evolution of triples during the pre-MS phase ($\tau$~$<$~5~Myr).  During this short timespan, the majority of our triple stars can undergo only one to a few KL oscillations.  We first incorporate quadrupole-level approximations, which adequately describe the majority of our triples during their short-lived pre-MS phase.  In \S2.2.2, we then consider octupole-level effects for the small fraction of our triples that can undergo several KL oscillations within $\tau$~$<$~5~Myr.   

The conjugate angular momenta of the inner and outer orbits with respect to their mean anomalies, i.e., assuming their orbits are circular, are:

\footnotesize
\begin{align}
 L_{\rm in,0} &= \frac{M_1 M_2}{M_1 + M_2}\sqrt{G (M_1 + M_2) a_{\rm in,0}} \nonumber \\
 L_{\rm out,0} &= \frac{M_3 (M_1 + M_2)}{M_1 + M_2 + M_3}\sqrt{G (M_1 + M_2 + M_3) a_{\rm out,0}} 
\end{align}
\normalsize

\noindent The true angular momenta of the inner and outer orbits are given by $J_x$ = $L_x(1-e_x^2)^{\nicefrac{1}{2}}$.  In our quadrupole-level approximations, the inner binary reaches a certain maximum eccentricity $e_{\rm in,max}$ after one KL timescale. In the absence of tidal effects (see \S2.3), the maximum eccentricity $e_{\rm in,max}$ satisfies:

\footnotesize
\begin{align}
 & 5\,{\rm cos}^2i_{\rm tot,0} - 3 + \frac{L_{\rm in,0}}{L_{\rm out,0}}\frac{{\rm cos}\,i_{\rm tot,0}}{\sqrt{1 - e_{\rm out,0}^2}}
 + \Big( \frac{L_{\rm in,0}}{L_{\rm out,0}} \Big)^2 \frac{e_{\rm in,max}^4}{1 - e_{\rm out,0}^2} \nonumber \\
 &+ e_{\rm in,\,max}^2 \Big[3 + 4 \frac{L_{\rm in,0}}{L_{\rm out,0}}\frac{{\rm cos}\,i_{\rm tot,0}}{\sqrt{1 - e_{\rm out,0}^2}}
 +  \Big( \frac{L_{\rm in,0}}{2 L_{\rm out,0}} \Big)^2 \frac{1}{1 - e_{\rm out,0}^2} \Big] = 0.
\end{align}
\normalsize

\noindent Our Eqn.~4 is identical to Eqn.~62 in \citet{Liu2015} and similar to Eqn.~A42 in \citet{Naoz2013} and Eqn.~24 in \citet{Anderson2017}.   At the quadrupole-level, only a limited range of initial mutual inclinations $i_{\rm tot,0}$ induce KL oscillations.  This ``KL window'' corresponds to where $e_{\rm in,max}$~$>$~0 is real (non-imaginary) according to Eqn.~4.  In the limit $L_{\rm in,0}$~$\ll$~$L_{\rm out,0}$, the KL window spans the canonical interval 39.2$^{\circ}$~$<$~$i_{\rm tot,0}$~$<$~140.8$^{\circ}$ and Eqn.~4 reduces to the classic equation:

\footnotesize
\begin{equation}
e_{\rm in,max} = \sqrt{1 - \frac{5}{3}\,{\rm cos}^2\,i_{\rm tot,0}}~~~.
\end{equation}
\normalsize

\noindent In the limit $L_{\rm in,0}$ $\ll$ $L_{\rm out,0}$, $e_{\rm max}$ approaches unity as the mutual inclination $i_{\rm tot,0}$~$\approx$~90$^{\circ}$ between the inner and outer orbits becomes orthogonal.  

In our baseline model, 57\% of our triples have $L_{\rm in,0}$/$L_{\rm out,0}$~$>$~0.1, and so their evolution differs slightly from the simple approximation in Eqn.~5. Moreover, our triples with larger $L_{\rm in,0}$/$L_{\rm out,0}$ tend to have shorter $\tau_{\rm KL}$. We therefore implement the more precise Eqn.~4 when evaluating the maximum eccentricity $e_{\rm in,out}$ achieved by the inner binary according to the quadrupole-level approximations. With increasing  $L_{\rm in,0}$/$L_{\rm out,0}$, $e_{\rm in,max}$\,$\rightarrow$\,1 occurs at larger inclinations  $i_{\rm tot,0}$~$>$~90$^{\circ}$.  For instance, for $L_{\rm in,0}$/$L_{\rm out,0}$~=~0.1 and $e_{\rm out}$~=~0.5,  $e_{\rm in,max}$\,$\rightarrow$\,1 is achieved at $i_{\rm tot,0}$~$\approx$~99$^{\circ}$.  Meanwhile, for $L_{\rm in,0}$/$L_{\rm out,0}$~=~1.0 and the same $e_{\rm out}$~=~0.5, $e_{\rm in,max}$\,$\rightarrow$\,1 occurs at $i_{\rm tot,0}$~$\approx$~125$^{\circ}$.

We investigate in detail the prescriptions for tidal evolution in \S2.3.  For a particular triple, the ability of the inner binary to tidally evolve toward smaller separations depends critically on the properties of the system.  On average, we find in \S2.3 that an inner binary will decay to short periods if the minimum periastron separation satisfies $r_{\rm peri,in}$~=~$a_{\rm in,0}$(1$-e_{\rm in,max}$)~$<$~5$R_1$.  In our quadrupole-level approximation, the inner binary evolves toward $e_{\rm in,max}$ given by Eqn.~4 after $\tau_{\rm KL}$, at which point we calculate the radius $R_1(\tau_{\rm KL})$ of the primary.  In Fig.~2, we show the cumulative fraction of systems with $r_{\rm peri,in}$~$<$~5$R_1$ as a function of age $\tau$.  At the quadrupole-level, 0.8\% of systems achieve $r_{\rm peri,in}$~$<$~5$R_1$ within $\tau$~$<$~5~Myr.  This fraction marginally increases to 1.0\% by 5~Gyr.

\begin{figure}[t!]
\centerline{
\includegraphics[trim=0.5cm 0.15cm 0.5cm 0.2cm, clip=true, width=3.6in]{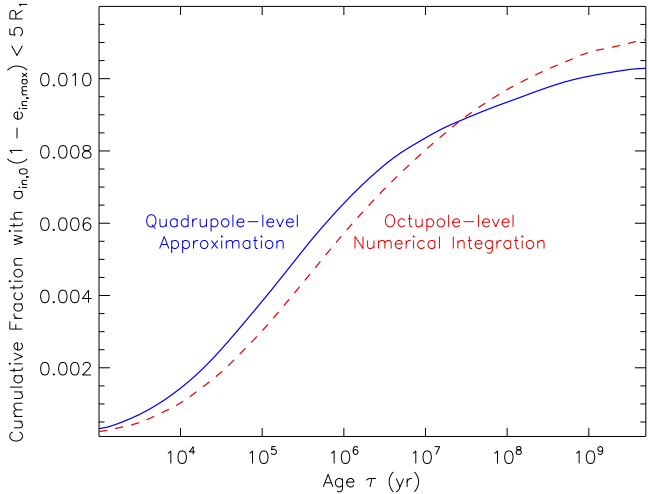}}
\caption{Cumulative fraction of systems that reach small periastron separations $r_{\rm peri,in}$ = $a_{\rm in,0}$(1$-e_{\rm in,max}$)~$<$~5$R_1$ due to KL oscillations in triples as a function of age~$\tau$ in our baseline model. We compare the quadrupole-level approximations (blue) to numerically integrated octupole-level secular equations (dashed red).   Approximately 0.8\% of systems (5\% of triples given $F_{\rm triple}$ = 0.15) have inner binaries that achieve $r_{\rm peri,in}$~$<$~5$R_1$ by $\tau$~=~5~Myr, the majority of which will tidally evolve toward shorter periods. Dynamical evolution of inner binaries toward small periastron separations $r_{\rm peri,in}$~$<$~5$R_1$ becomes less efficient beyond $\tau$~$>$~10~Myr, not only because $R_1$ is smaller, but also because systems with longer $\tau_{\rm KL}$ are less likely to evolve toward small $r_{\rm peri,in}$ (see text for discussion).}
\end{figure}

\subsubsection{Octupole-level Effects}

We now consider octupole-level effects for the triples that undergo several KL oscillations during the pre-MS phase.  The relative strength of the octupole level with respect to the quadrupole level is:

\footnotesize
\begin{equation}
\epsilon_{\rm oct} = \frac{M_1 - M_2}{M_1+M_2} \frac{a_{\rm in}}{a_{\rm out}} \frac{e_{\rm out}}{1 - e_{\rm out}^2} . 
\end{equation}
\normalsize

\noindent In our baseline model, 18\% of our triples have $\epsilon_{\rm oct}$~$>$~0.01 such that octupole-level effects can potentially become important.    Moreover, the systems with larger $\epsilon_{\rm oct}$ tend to have shorter KL timescales $\tau_{\rm KL}$.  In particular, 9\% of the triples in our baseline model have both $\epsilon_{\rm oct}$~$>$~0.01 and $\tau_{\rm KL}$~$<$~0.5~Myr, and can therefore undergo many KL oscillations modulated by octupole-level effects during the pre-MS phase.

We calculate the octupole-level secular evolution of triples according to the equations provided in Appendix~A of \citet{Liu2015}, which are similar to the equations presented in Appendix B of \citet{Naoz2013}. Specifically, we numerically integrate the differential equations A1-A4 and A6-A7 in \citet{Liu2015}.  We do not calculate the changes in the longitudes of ascending nodes (their Eqn.~A5), which are related to the arguments of periastron $\omega_{\rm in}$ and $\omega_{\rm out}$ and therefore not needed to numerically integrate the system.  For the remaining equations in Appendix~A of \citet{Liu2015}, the right-hand sides all depend on previously defined parameters of the system: $i_{\rm tot}$, $e_{\rm in}$, $e_{\rm out}$, $\omega_{\rm in}$, $\omega_{\rm out}$, $L_{\rm in}$, $L_{\rm out}$, $\tau_{\rm KL}$, and $\epsilon_{\rm oct}$.  In their Eqns.~A3 and A4, \citet{Liu2015} provide the differential equations $di_{\rm in}/dt$ and $di_{\rm out}/dt$ for the inclinations of the inner and outer orbits with respect to the total angular momentum vector, respectively.  We use the relation $i_{\rm tot}$~=~$i_{\rm in}$~+~$i_{\rm out}$ to evaluate the differential equation $di_{\rm tot}/dt$~=~$di_{\rm in}/dt$~+~$di_{\rm out}/dt$ for the total mutual inclination between the inner and outer orbits.  We utilize a Runge-Kutta technique with adaptive time steps to numerically integrate the secular equations for each triple.

In Fig.~3, we display the inner binary's eccentricity evolution for a particular triple where octupole-level effects are important in the context of forming close binaries during the pre-MS phase. This specific system initially has $M_1$~=~1\Msun, $M_2$~=~$M_3$~=~0.4\Msun, $a_{\rm in,0}$~=~30~AU, $a_{\rm out,0}$~=~300~AU, $e_{\rm in,0}$~=~0.01, $e_{\rm out,0}$~=~0.5, $\omega_{\rm in,0}$~=~$\omega_{\rm out,0}$~=~0$^{\circ}$, and $i_{\rm tot,0}$~=~104$^{\circ}$, which provides $\tau_{\rm KL}$~=~0.054~Myr and $\epsilon_{\rm oct}$ = 0.029.  While the KL timescale according to Eqn.~2 is $\tau_{\rm KL}$~=~0.054~Myr, the actual period of the KL cycles determined by numerically integrating the secular differential equations is $\approx$0.2 Myr.  The time for the inner binary to first approach $e_{\rm in,max}$ depends on the initial properties of the system. On average, we find it to be several times $\tau_{\rm KL}$.

During the first KL cycle of our example triple in Fig.~3, the inner binary reaches $e_{\rm in,max}$~=~0.9935, which is similar to the quadrupole-level approximation $e_{\rm in,max}$~=~0.9924 given by Eqn.~4.  The minimum periastron distance during the first KL oscillation is $r_{\rm peri,in}$~=~42\Rsun.  This is still $\approx$11 times larger than the $R_1$ = 3.9\Rsun\ pre-MS primary at the time of closest approach, and so we do not expect tidal dissipation to effectively bring the inner binary to closer separations.  During the third KL oscillation, however, the inner binary achieves a much larger eccentricity $e_{\rm in,max}$~=~0.9994.  The periastron separation $r_{\rm peri,in}$ = 4.1\Rsun\ is substantially smaller, and the pre-MS primary has only contracted slightly to $R_1$~=~2.6\Rsun.  Given $r_{\rm peri,in}$/$R_1$~=~1.6, the primary would overfill its Roche lobe at periastron in the absence of tidal effects.  In reality, prior to reaching such small periastron separations, tidal energy dissipation will cause the inner binary to decay toward shorter orbital periods (see \S2.3).  We note the companion masses, the initial inner and outer separations, and the eccentricity of the outer orbit in our example triple are close to the means of their corresponding probability distribution functions in our baseline model. Our octupole-level example is therefore representative of a non-trivial fraction of triples in our simulations.

\begin{figure}[t!]
\centerline{
\includegraphics[trim=0.5cm 0.15cm 0.6cm 0.2cm, clip=true, width=3.7in]{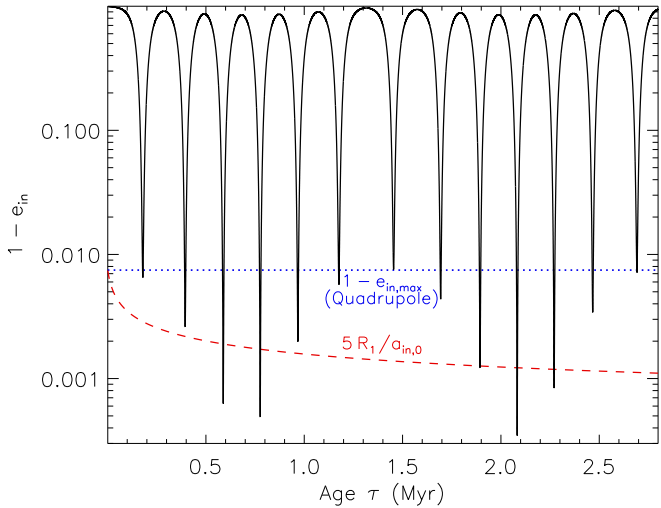}}
\caption{Octupole-level eccentricity evolution (black) of an inner binary in a triple with $M_1$~=~1\Msun, $M_2$~=~$M_3$~=~0.4\Msun, $a_{\rm in,0}$~=~30~AU, $a_{\rm out,0}$~=~300~AU, $e_{\rm in,0}$~=~0.01, $e_{\rm out,0}$~=~0.5, $\omega_{\rm in,0}$~=~$\omega_{\rm out,0}$~=~0$^{\circ}$, and $i_{\rm tot,0}$~=~104$^{\circ}$. We also show the quadrupole-level approximation for the maximum eccentricity $e_{\rm in,max}$ = 0.9924 (dotted blue) and the decrease in the ratio 5$R_1$/$a_{\rm in,0}$ due to the contraction of the pre-MS primary (dashed red).  During the first KL oscillation, the inner binary reaches a minimum periastron separation $r_{\rm peri,in}$~=~11$R_1$~=~42\Rsun.   Because of large octupole-level effects $\epsilon_{\rm oct}$~=~0.029, the inner binary approaches a much smaller separation $r_{\rm peri,in}$~=~1.6$R_1$~=~4\Rsun\ during its third KL cycle, and so will tidally evolve toward smaller separations and eccentricities (not included in this simulation - see Fig.~4).}
\end{figure}

We numerically integrate the octupole-level secular equations for our baseline population of triples.  We evolve each triple for 5~Gyr, for 200$\tau_{\rm KL}$, or until $r_{\rm peri,in}$~$<$~5$R_1$, whichever occurs first.  As done in \S2.2.1 with our quadrupole-level approximations, we show in Fig.~2 the cumulative fraction of systems with $r_{\rm peri,in}$~=~$<$~5$R_1$ as a function of age based on the octupole-level simulations.  At younger ages $\tau$~$<$~10~Myr, numerical integration of the octupole-level secular equations yields a fraction of systems with $r_{\rm peri,in}$~=~$<$~5$R_1$ that is smaller than the fraction determined in our quadrupole-level approximation.  This is because the inner binary actually takes several times $\tau_{\rm KL}$, on average, to first approach $e_{\rm in,max}$.  At older ages $\tau$~$>$~10~Myr, the fraction of systems with $r_{\rm peri,in}$~=~$<$~5$R_1$ in the octupole-level simulation is slightly larger than the quadrupole-level approximation.  As expected, octupole-level effects widen the parameter space of triples that can achieve very large eccentricities, albeit mildly. For both the quadrupole-level and octupole-level simulations, we find that $\approx$0.8\% of systems achieve $r_{\rm peri,in}$~=~$<$~5$R_1$ by 5~Myr.  

We find the rate of forming close binaries via secular evolution in triples diminishes with time beyond $\tau$~$>$~10~Myr due to four compounding effects.  First, half of the triples in our baseline model have $\tau_{\rm KL}$~$<$~10~Myr.  To quadrupole level and assuming tides effectively decay the orbits of the inner binaries that achieve $r_{\rm peri,in}$~$<$~5$R_1$ within a single or a few KL cycles (see \S2.3), the close binary fraction can at most double beyond $\tau$~$>$~10~Myr.  Second, even octupole-level effects dominate at early times $\tau$~$<$~10~Myr because systems with larger octupole strengths generally have shorter KL timescales.  For instance, of the triples with $\epsilon_{\rm oct}$~$>$~0.01 in our baseline model, 87\% have $\tau_{\rm KL}$~$<$~10~Myr.  Third, the radius $R_1$~$\approx$~1\Rsun\ of the primary beyond $\tau$~$>$~10~Myr is substantially smaller than its size $R_1$~$\approx$~2\,-\,10 \Rsun\ during the early pre-MS phase.  An inner binary with a smaller MS primary must dynamically evolve to a much larger $e_{\rm in,max}$, i.e., a smaller $r_{\rm peri,in}$~$\lesssim$~5$R_1$, for tidal effects to become important.  Finally, triples with longer $\tau_{\rm KL}$ have systematically wider $a_{\rm in,0}$.  In our baseline model, the median inner binary separation of triples with $\tau_{\rm KL}$~$<$~0.1~Myr is $a_{\rm in,0}$~=~11~AU, while for $\tau_{\rm KL}$~$=$~5\,-\,100~Myr, the median inner binary separation is $a_{\rm in,0}$~=~21~AU. To reach the same $r_{\rm peri,in}$ = $a_{\rm in,0}$(1$-e_{\rm in,max})$, systems with wider $a_{\rm in,0}$ must evolve toward larger $e_{\rm in,max}$.  These four effects lead to the flattening of the curves beyond $\tau$~$>$~10~Myr in Fig.~2.

\subsection{Tidal Evolution}

\subsubsection{Types and Application of Tidal Effects}

We next incorporate tidal effects in the dynamical evolution of the inner binaries.  There are two mechanisms by which tides can alter the orbital parameters.  First, tidal precession can cause the secular evolution of the inner binary to decouple from the KL oscillation, thereby preventing the inner binary from reaching as large an eccentricity as it would have achieved in the absence of tidal effects \citep{Fabrycky2007,Liu2015,Anderson2017}.  In this scenario, the energy of the system is conserved, and so the separation $a_{\rm in}$ of the inner binary remains constant.  Second, energy of the inner orbit is tidally dissipated into the interior of the stellar components and then subsequently radiated \citep{Zahn1977,Hut1981,Kiseleva1998,Eggleton1998,Eggleton2001,Fabrycky2007}.  This process is commonly known as tidal friction, and so we use the terms tidal friction and tidal energy dissipation interchangeably.  Because energy is lost from the system, both the eccentricity $e_{\rm in}$ and separation $a_{\rm in}$ of the inner binary evolve toward smaller values.

\citet{Liu2015} and \citet{Anderson2017} investigated the effects of precession in the context of secular evolution of triple stars.   They determined that general relativistic precession can limit the maximum eccentricity of inner binaries that are initially close, precession due to the tidal bulge can limit the maximum eccentricity of inner binaries that are initially wide, and precession due to rotation-induced oblateness is important only if the stars are rapidly rotating. In their analysis, \citet{Liu2015} and \citet{Anderson2017} purposefully excluded the effects of tidal friction and tidal disruption so that they could focus strictly on the role of precession. By ignoring these effects, however, inner binaries in their models were allowed to evolve toward very small periastron separations.  For example, \citet{Anderson2017} examined the secular evolution of a triple with an inner binary parameterized by $M_1$ = 1.0\Msun, $M_2$ = 0.5\Msun, and $P_{\rm in}$~=~15~days ($a_{\rm in}$~=~29\Rsun; see their Fig.~9). For a broad range of mutual inclinations, the inner binary approached a maximum eccentricity $e_{\rm in,lim}$~=~0.93 limited by tidal and general relativistic precession.  This limiting eccentricity corresponds to a periastron separation of $r_{\rm peri,in}$~=~2.1\Rsun, which is inside the tidal disruption separation of $r_{\rm peri,in}$~$\approx$~2.5\,$R_1$. Similarly, \citet{Liu2015} modeled a triple with $M_1$~=~1.0\Msun, $R_1$~=~1.0\Rsun, $M_2$~=~0.5\Msun, $R_2$~=~0.5\Rsun, and $a_{\rm in,0}$~=~1~AU (see their Fig.~12 and Tables~3\,-\,4).  They measured a limiting eccentricity of $e_{\rm in,lim}$ = 0.9982, corresponding to a periastron separation of $r_{\rm peri,in}$ =~0.4\Rsun\ that is actually inside the radii of the stellar components. Most inner binaries will be affected by tidal energy dissipation ($r_{\rm peri,in}$~$\approx$~2.5\,-\,5\,$R_1$) and possibly become tidally disrupted ($r_{\rm peri,in}$~$\lesssim$~2.5\,$R_1$) long before precession can limit the eccentricity.  We further demonstrate in \S2.3.5 that for our population of triples in particular, which have initially much wider inner binaries than that considered in previous studies, tidal precession is negligible.  In our full numerical simulations, we therefore consider only tidal friction.

For solar-type MS and pre-MS primaries in binaries with small to moderate eccentricities, tidal friction is most efficient through eddy viscosity in the convective stellar envelopes as parameterized by the weak-friction equilibrium-tide model \citep{Zahn1977,Hut1981,Kiseleva1998,Eggleton1998}.  In the weak-friction equilibrium-tide model, the companion raises and sustains a tidal bulge on the primary star throughout its entire orbit.  If the orbit is eccentric or not synchronized with the rotation rate of the primary, there is a phase lag between the primary's bulge and companion.  This phase lag yields a net torque that causes both orbital energy and angular momentum to be transferred from the orbit to the primary's bulge.  Convection within the primary's envelope subsequently deposits the transferred angular momentum and energy into the stellar interior on the eddy turnover timescale.  While the transferred angular momentum spins up the primary, the transferred energy is ultimately lost from the system via radiation.

 For our triples with $\langle a_{\rm in,0} \rangle$~$\sim$~30~AU, however, the inner binaries must generally reach very large eccentricities $e_{\rm in,max}$~$\gtrsim$~0.99 in order for tides to effectively decay the orbits.  At such large eccentricities, tidal energy dissipation occurs solely near periastron, and so tidal evolution averaged over the entire orbit as parameterized by the weak-friction equilibrium-tide model in \citet{Zahn1977}, \citet{Hut1981}, and \citet{Eggleton1998} is no longer applicable.  In this nearly parabolic regime, the companion raises a large tidal bulge on the primary near periastron, and the energy associated with this tidal deformation is subsequently dissipated into the primary via non-radial dynamical oscillations \citep{Fabian1975,Press1977}.  We note that \citet{Fabian1975} and \citet{Press1977} originally discussed this mechanism in the context of tidally capturing two unbound stars in an initially hyperbolic orbit into a stable, gravitationally bound binary.  This same process may occur in a highly eccentric binary that passes within a few stellar radii at periastron but spends most of its orbit at large separations where tidal effects are negligible.  The tidal evolution of our inner binaries as they approach nearly parabolic orbits in their KL cycles is therefore similar to the subsequent evolution of tidally captured binaries \citep[e.g., see][]{McMillan1986,Kochanek1992,Mardling1995}. In the present study, we sometimes refer to the inner binary's rapid evolution toward smaller separations due to dynamical tides as ``tidal capture'', even though the inner binaries in our simulations are always gravitationally bound. 

Once the eccentricity of the inner binary is reduced via dynamical tides and the primary's rotation rate begins to pseudo-synchronize with the angular frequency of the inner binary near periastron, energy dissipation by raising a new tidal bulge at each periastron passage is suppressed.  \citet{Mardling1995} showed that this transition between the large-eccentricity dynamical regime to the small-eccentricity regime where only long-term dissipative processes circularize the orbit occurs near $e$~$\approx$~0.8.  Observations of solar-type binaries with intermediate periods exhibit a deficit of binaries with $e$~$>$~0.8 compared to systems with $e$~=~0.6\,-\,0.8 \citep[][references therein]{Duquennoy1991,Raghavan2010,Moe2017}.  This indicates that binaries that may initially have $e$~$\gtrsim$~0.8 quickly evolve toward $e$~$<$~0.8, at which point weak-friction equilibrium tides circularize and decay the orbit on much longer timescales.  In our baseline model, we follow the parameterizations for rapid tidal energy dissipation via non-radial dynamical modes as treated in \citet{Press1977} when $e_{\rm in}$~$>$~$e_{\rm trans}$~=~0.8, and subsequently follow the prescriptions for weak-friction equilibrium tides as developed by \citet{Zahn1977}, \citet{Hut1981}, and \citet{Eggleton1998} when $e_{\rm in}$~$<$~$e_{\rm trans}$~=~0.8.  We also consider a smaller transition eccentricity $e_{\rm trans}$~=~0.5 (Model~A9) to determine the importance of the transition in the dominant tidal mechanism. 

\subsubsection{Prescriptions for Dynamical Tides}

The energy associated with raising a tidal bulge on the primary at periastron and subsequent dissipation of that energy via non-radial dynamical oscillations is \citep{Press1977}:

\footnotesize
\begin{equation}
 \Delta E_{\rm tide} = \frac{G M_2^2}{R_1} \Big( \frac{r_{\rm peri,in}}{R_1} \Big)^{-6} T_2(\eta) , 
\end{equation}
\normalsize

\noindent where:

\footnotesize
\begin{equation}
 \eta = \Big( \frac{M_1}{M_1 + M_2} \Big)^{1/2} \Big( \frac{r_{\rm peri,in}}{R_1} \Big)^{3/2}.
\end{equation}
\normalsize

\noindent and $T_2$ is the efficiency of energy deposition via the dominant $l$=2 quadrupole tide. The efficiency $T_2$ not only depends on $\eta$, but also on the interior structure and elasticity of the primary star.  For a $n$~=~3 polytrope, which is representative of a solar-type MS star, \citet{Press1977} calculate $T_2$~$\approx$~0.2 for $\eta$~=~1 and $T_2$~$\approx$~0.003 for $\eta$~=~10.  Parameterizing the relation as a simple power-law $T_2$~$\approx$~0.3$\eta^{-2}$, we find:

\footnotesize
\begin{equation}
 \Delta E_{\rm tide} = f_{\rm dyn} \frac{M_1+M_2}{M_1} \frac{G M_2^2}{R_1} \Big( \frac{r_{\rm peri,in}}{R_1} \Big)^{-9} , 
\end{equation}
\normalsize

\noindent where $f_{\rm dyn}$~=~0.3 for MS primaries modeled by $n$~=~3 polytropes.  For $n$~=~3/2 polytropes, which more closely resembles the density profiles of fully convective pre-MS stars, \citet{McMillan1986} estimates the energy to be 1.84$^2$~$\approx$~3.4 times larger.  In our baseline model, we use Eqn.~9  with $f_{\rm dyn}$~=~0.3 for MS primaries with ages $\tau$~$>$~30~Myr, $f_{\rm dyn}$~=~1.0 for fully convective pre-MS primaries with $\tau$~$<$~3~Myr, and interpolate $f_{\rm dyn}$ with respect to log~$\tau$ between these two regimes.  More recent calculations show the efficiency $T_2$ of transferring energy via the quadrupole dynamical tide is $\approx$\,3\,-\,10 times smaller than the original estimate by \citet{Press1977}, depending on $\eta$ \citep{McMillan1986,Mardling1995,Lai1997}.  We therefore consider an additional simulation with $T_2$ = 0.03$\eta^{-2}$ (Model A10).  In this model, we use Eqn.~9 with $f_{\rm dyn}$~=~0.03 for MS primaries, $f_{\rm dyn}$~=~0.1 for pre-MS primaries, and interpolate with respect to age as done in our baseline model. 

In the high-eccentricity regime, the energy $\Delta$E$_{\rm tide}$ given by Eqn.~9 is transferred from the orbital energy $E_{\rm orb,in}$~=~$-$G$M_1 M_2$/(2$a_{\rm in}$) into the stellar interiors during each periastron passage. Due to the impulse of tidal energy exchange at periastron, the periastron separation remains relatively constant.  As $E_{\rm orb,in}$ becomes more negative and $a_{\rm in}$ decreases, $e_{\rm in}$ also decreases to maintain the same $r_{\rm peri,in}$.  If $r_{\rm peri,in}/R_1$ is too large and/or the inner binary evolves too quickly back toward large periastron separations in its KL cycle, the inner binary will not tidally decay to smaller separations.  If instead sufficient energy is transferred over multiple $N_{\rm in,orbit}$ periastron passages, i.e., $N_{\rm in,orbit}\times\Delta E_{\rm tide}$~$\sim$~$|E_{\rm orb,in}|$, then the inner binary will tidally decay toward smaller separations and eccentricities. In our simulations, we numerically integrate the octupole-level secular equations while decreasing the orbital energy of the inner binary according to Eqn.~9 during each orbit, i.e.,:

\begin{equation}
\Big( \frac{da_{\rm in}}{dt} \Big)_{\rm tide} = \frac{a_{\rm in}}{P_{\rm in}} \frac{\Delta E_{\rm tide}}{E_{\rm orb,in}}
\end{equation}

There is an inherent inconsistency in the inclusion of a dynamical tide in our secular (orbit-averaged) model.  For example, based on our analytic prescriptions for dynamical tides, the inner binaries decay toward smaller eccentricities $e_{\rm in}$~=~$e_{\rm trans}$~$\approx$~0.8 monotonically.  In full numerical simulations, however, tidally captured binaries exhibit slightly chaotic behavior in which the eccentricities actually evolve non-monotonically \citep{Mardling1995}.  For our population synthesis models, it is prohibitive to numerically integrate the dynamical oscillations and tidal energy dissipation in all of our systems. Fortunately, the simulations of tidally captured binaries show a long-term average behavior whereby the binaries decay toward smaller eccentricities $e_{\rm in}$~$=$~$e_{\rm trans}$~$\approx$~0.8 on timescales of a few hundred to few thousand orbits while nearly preserving $r_{\rm peri, in}$~$\approx$~(3\,-\,5)$R_1$ \citep{McMillan1986, Kochanek1992, Mardling1995}.  Our analytic prescriptions for dynamical tides are consistent with the long-term behavior and timescales of these numerical simulations.

The primary reason that dynamical tides are much more important in our models than in previous studies is the nature of our initial inner binary population.  All of our inner binaries initially have $a_{\rm in,0}$~$>$~0.5~AU, the majority of which have $a_{\rm in,0}$~=~10\,-\,100~AU.  In order to achieve $P_{\rm in}$~$<$~10~days, the inner binaries must first pass through the high-eccentricity regime $e_{\rm in}$~$>$~$e_{\rm trans}$~$\approx$~0.8 where dynamical tides are relevant.  If dynamical tides were ignored, the simulated close binary fraction would be reduced, especially during the pre-MS phase.  Without dynamical tides, some inner binaries would reach very small periastron separations $r_{\rm peri,in}$~$\lesssim$~2.5$R_1$ and become tidally disrupted.  For instance, our example system in Fig.~3 achieves $r_{\rm peri,in}$~$=$~1.6$R_1$ in the absence of tidal effects.  In other cases, the pre-MS primaries would contract before weak-friction equilibrium tides could effectively decay the inner orbits.  For many of our systems that form close binaries via dynamical tides, weak-friction equilibrium tides alone could also decay the orbits toward shorter separations, but only after hundreds to thousands of KL oscillations well beyond the pre-MS phase.  In order to form close binaries ($a_{\rm in}$~$<$~0.1~AU) from initially much wider binaries ($a_{\rm in,0}$~$\approx$~10\,-\,100 AU) via KL cycles during the pre-MS phase, dynamical tides must be included. 

\subsubsection{Prescriptions for Equilibrium Tides}

Once the eccentricity of the inner binary decays to $e_{\rm in}$~$<$~$e_{\rm trans}$~=~0.8 via dynamical tides, the primary has become pseudo-synchronized with the orbital angular frequency near periastron and energy dissipation proceeds solely through  weak-friction equilibrium tides.  We utilize prescriptions for weak-friction equilibrium tides as formalized by \citet{Zahn1977}, \citet{Hut1981}, and \citet{Eggleton1998}, and summarized in \citet{Hurley2002} and \citet{Belczynski2008}.  The tidal equations for the inner binary's separation and eccentricity evolution are:

\footnotesize
\begin{align}
\Big( \frac{da_{\rm in}}{dt} \Big)_{\rm tide} = & -6 F_{\rm tide} \frac{k_1}{\tau_{\rm conv,1}}
         q_2 (1 + q_2) \Big( \frac{R_1}{a_{\rm in}} \Big)^8 \frac{a_{\rm in}}{(1 - e_{\rm in}^2)^{\nicefrac{15}{2}}} \nonumber \\
         & \Big[ f_1 (e_{\rm in}^2) - (1 - e_{\rm in}^2)^{\nicefrac{3}{2}} f_2 (e_{\rm in}^2)\frac{\Omega_1}{\Omega_{\rm in}} \Big],
\end{align}

\begin{align}
\Big( \frac{de_{\rm in}}{dt} \Big)_{\rm tide} = & -27 F_{\rm tide} \frac{k_1}{\tau_{\rm conv,1}}
         q_2 (1 + q_2) \Big( \frac{R_1}{a_{\rm in}} \Big)^8 \frac{e_{\rm in}}{(1 - e_{\rm in}^2)^{\nicefrac{13}{2}}} \nonumber \\
         & \Big[ f_3 (e_{\rm in}^2) - \frac{11}{18}(1 - e_{\rm in}^2)^{\nicefrac{3}{2}} f_4 (e_{\rm in}^2)\frac{\Omega_1}{\Omega_{\rm in}} \Big],
\end{align}
\normalsize

\noindent where $F_{\rm tide}$ is a scaling factor, $k_1$ is the apsidal motion constant of the primary (which is half its tidal Love number), $\tau_{\rm conv,1}$ is the timescale for tidal energy dissipation within the envelope of the primary star via convection, $\Omega_1$ is the angular rotational frequency of the primary star, $\Omega_{\rm in}$~=~[G($M_1$+$M_2$)/$a_{\rm in}^3$]$^{\nicefrac{1}{2}}$ is the mean angular frequency of the inner binary, and the eccentricity functions are:

\footnotesize
\begin{align}
f_1(e_{\rm in}^2) & = 1 + (31/2)e_{\rm in}^2 + (255/8)e_{\rm in}^4 + (185/16)e_{\rm in}^6 + (25/64)e_{\rm in}^8, \nonumber \\
f_2(e_{\rm in}^2) & = 1 + (15/2)e_{\rm in}^2 + (45/8)e_{\rm in}^4 + (5/16)e_{\rm in}^6, \nonumber \\
f_3(e_{\rm in}^2) & = 1 + (15/4)e_{\rm in}^2 + (15/8)e_{\rm in}^4 + (5/64)e_{\rm in}^6,~{\rm and} \nonumber \\
f_4(e_{\rm in}^2) & = 1 + (3/2)e_{\rm in}^2 + (1/8)e_{\rm in}^4.
\end{align}
\normalsize

For stellar-mass companions, the timescales for spin-orbit alignment and pseudo-synchronization are orders of magnitude shorter than the timescales for tidal circularization \citep{Zahn1977,Hut1981}.  The primaries in several highly eccentric binaries are observed to be rotating close to pseudo-synchronization \citep{Hall1986,Zimmerman2017}.  In our models, primaries first achieve spin-orbit alignment and pseudo-synchronization when the inner binaries evolve from $e_{\rm in}$~$\gtrsim$~0.99 to $e_{\rm in}$~=~$e_{\rm trans}$~=~0.8 via dynamical tides, regardless of the their original spin momentum vectors.  Pseudo-synchronization is given by \citet{Hut1981}:

\footnotesize
\begin{equation}
 \frac{\Omega_1}{\Omega_{\rm in}} = 1 + 6e_{\rm in}^2 + (3/8)e_{\rm in}^4 + (223/8)e_{\rm in}^6,
\end{equation}  
\normalsize

\noindent so that $\Omega_1$/$\Omega_{\rm in}$~$\approx$~12 when $e_{\rm in}$~=~0.8 and the ratio $\Omega_1$/$\Omega_{\rm in}$~$\approx$~1.0\,-\,1.5 approaches unity as $e_{\rm in}$~$<$~0.3 becomes small.   For inner binaries that decay via weak-friction equilibrium tides, we set $\Omega_1$ according to Eqn.~14 so that the primary remains pseudo-synchronized with the eccentric orbit of the inner binary. 

For a $M_1$~=~1\Msun\ MS primary, the apsidal motion constant is close to $k_1$~$\approx$~0.02 \citep{Claret2004}.  For a large, convective 1\Msun\ pre-MS primary with log~g (cm\,s$^{-2}$)~$<$~4.0, i.e., $\tau$~$<$~3~Myr according to our adopted Dartmouth pre-MS tracks, the apsidal motion constant $k_1$~$\approx$~0.13 is larger \citep{Claret2012}.  For our simulations, we adopt $k_1$ = 0.13 for $\tau$~$<$~3~Myr, $k_1$~=~0.02 for ages $\tau$~$>$~30~Myr beyond the zero-age MS, and we interpolate with respect to log $\tau$ between these two regimes.  

The convective eddy turnover timescale of the primary is \citep{Rasio1996,Hurley2002}:

\footnotesize
\begin{equation}
 \tau_{\rm conv,1} = \Big[ \frac{M_{\rm 1,env}R_{\rm 1,env}(R_1 - R_{\rm 1,env}/2)}{3 L_1} \Big]^{\nicefrac{1}{3}}
\end{equation}
\normalsize

\noindent where $M_{\rm 1,env}$ and $R_{\rm 1,env}$ are the mass and depth of the convective envelope, respectively.  For a very young $\tau$~=~10$^3$~yr pre-MS primary that is fully convective, the parameters are $M_{\rm 1,env}$~$\approx$~$M_1$~=~1\Msun, $R_{\rm 1,env}$~$\approx$~$R_1$~=~10\Rsun, and $L_1$~=~30\Lsun, so that $\tau_{\rm conv,1}$~=~0.35~yr.  For a solar-type MS star, the envelope quantities are $R_{\rm 1,env}$~$\approx$~0.3\Rsun\ and $M_{\rm 1,env}$~=~0.02\Msun\ \citep{Dalsgaard1991}, which provides a shorter convection timescale $\tau_{\rm conv,1}$~=~0.05~yr.  In our simulations, we adopt $\tau_{\rm conv,1}$~=~0.05~yr for MS primaries with $\tau$~$>$~30~Myr, and linearly interpolate between $\tau_{\rm conv,1}$~=~0.35~yr at $\tau$ = 10$^3$~yr and $\tau_{\rm conv,1}$~=~0.05~yr at $\tau$~=~30~Myr with respect to log~$\tau$.  

Our Eqns.~11\,-\,12 incorporate only tidal friction dissipated within the primary star.  Because the tidal equations depend very strongly on stellar radius, tidal energy dissipation within the secondary star becomes important only when $R_2$~$\gtrsim$~0.8$R_1$, i.e., $q_2$~$\gtrsim$~0.8 in our case of coeval pre-MS and MS stellar binaries.  We account for the slight enhancement in tidal energy dissipation when $q_2$~$>$~0.8, reaching twice the values provided in Eqns.~11\,-\,12 when $q_2$~=~1.

Given the above relations, we find that a scaling factor of $F_{\rm tide}$~=~20 best reproduces the observed circularization period of stellar populations as a function of age (see \S3).  \citet{Meibom2005} and \citet{Belczynski2008} also find that the simplified theory of tidal dissipation as parameterized above slightly underestimates the true tidal strengths as constrained by observations. We set $F_{\rm tide}$~=~20 in our baseline model, but also consider the nominal value $F_{\rm tide}$~=~1 (Model~A11).

\subsubsection{Example of Tidal Evolution}

We numerically integrate the differential equations governing tidal dissipation above in parallel with the octupole-level secular equations discussed in \S2.2. We illustrate the nature of our tidal model in Fig.~4, where we present the dynamical evolution of the same triple analyzed in Fig.~3 and \S2.2 while incorporating dynamical tides at large eccentricities and weak-friction equilibrium tides at small eccentricities.  As expected, tides are negligible during the first KL oscillation when the closest periastron separation is $r_{\rm peri,in}$~=~11$R_1$.  During the second KL cycle, the inner binary reaches a smaller periastron separation of $r_{\rm peri,in}$~=~6.3$R_1$ and tidally decays slightly from $a_{\rm in}$~=~30~AU to $a_{\rm in}$~=~26~AU.  However, the inner binary evolves back toward small eccentricities and large periastron separations in its KL oscillation.  Note that if the inner binary had reached $r_{\rm peri,in}$~=~5$R_1$, i.e., our crude criterion for evolving toward small separations in \S2.2, the energy dissipation would have been an order of magnitude larger according to Eqn.~9.  During the third KL cycle, the inner binary approaches a very small periastron separation $r_{\rm peri,in}$~=~3$R_1$ and tidally evolves to $e_{\rm in}$~=~$e_{\rm trans}$~=~0.8, corresponding to $a_{\rm in}$~=~0.15~AU, after $\approx$1,300 orbits.  Subsequent energy dissipation via  weak-friction equilibrium tides decays the orbit further to $a_{\rm in}$~=~12\,\Rsun~=~0.055~AU ($P_{\rm in}$~=~3.9~days) and $e_{\rm in}$ = 0.01 after an additional $\Delta \tau$~=~0.1~Myr.  We emphasize that if either (1) octupole-level effects were neglected, (2) tidal energy dissipation via non-radial dynamical oscillations at large eccentricities were not included, or (3) the larger tidal radius of the pre-MS primary was not considered, then this particular triple would not have produced a close binary. By incorporating all three of these effects, we show that an inner binary born with a wide separation $a_{\rm in,0}$~=~30~AU can migrate to a very small separation $a_{\rm in}$~=~0.055~AU due to KL oscillations and tidal friction during the pre-MS phase.

\begin{figure}[t!]
\centerline{
\includegraphics[trim=0.4cm 0.35cm 1.3cm 0.2cm, clip=true, width=3.7in]{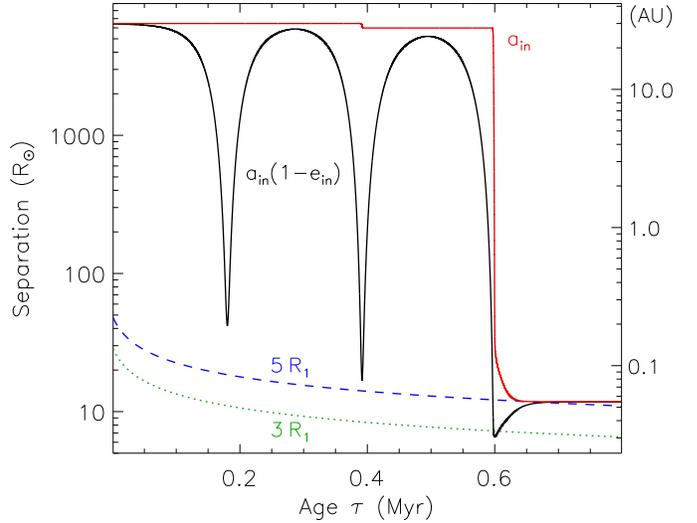}}
\caption{Separation evolution of the inner binary in the same triple as analyzed in Fig.~3, but where we include the effects of tidal energy dissipation via dynamical oscillations at large eccentricities $e_{\rm in}$~$>$~0.8 and weak-friction equilibrium tides at small eccentricities $e_{\rm in}$~$<$~0.8.  We display the periastron separation $r_{\rm peri,in}$~=~$a_{\rm in}$(1$-e_{\rm in}$) (black), semi-major axis separation $a_{\rm in}$ (red), and the contraction of the pre-MS primary as indicated by 3$R_1$ (dotted green) and 5$R_1$ (dashed blue).  Previous studies that incorporate only weak-friction equilibrium tides generally require hundreds of KL oscillations for tidal friction to be effective.  We instead find that dynamical tides at large eccentricities and small periastron separations $r_{\rm peri,in}$~$\approx$~(3\,-\,5)$R_1$, as motivated by numerical simulations of tidally captured binaries, is orders of magnitude more efficient.  For this example, the inner binary tidally decays from $a_{\rm in}$~$\approx$~26~AU to $a_{\rm in}$~$\approx$~0.15~AU via dynamical oscillations during its third KL cycle.  The inner binary subsequently circularizes via weak-friction  equilibrium tides after an additional $\Delta \tau$~=~0.1~Myr, settling into an orbit with $a_{\rm in}$~=~12\,\Rsun~=~0.055~AU ($P_{\rm in}$~=~3.9~days).}
\end{figure}

 The equations for weak-friction equilibrium tides preserve the same orbital angular momentum $J_{\rm in}$~$\propto$~[$a_{\rm in}$(1$-e_{\rm in}^2$)]$^{\nicefrac{1}{2}}$ assuming the primary's spin angular momentum $J_1$~$\ll$~$J_{\rm in}$ is negligible.  In the high-eccentricity regime when $e_{\rm in}$~$>$~0.8, the orbital angular momentum $J_{\rm in}$ can become small and so the spin angular momentum $J_1$ is no longer negligible. $J_{\rm in}$ is therefore not conserved as the inner binary is initially tidally captured toward smaller separations and the primary first becomes pseudo-synchronized with the eccentric orbit. Our treatment of tidal friction in the high-eccentricity regime instead preserves the periastron separation $r_{\rm peri,in}$~=~$a_{\rm in}$(1$-e_{\rm in}$). Numerical simulations of truly tidally captured binaries show their periastron separations $r_{\rm peri}$ remain relatively constant as the system evolves via dynamical tides from hyperbolic configurations down to $e$~$\approx$~0.6\,-\,0.8 on timescales of a few hundred to a few thousand orbits \citep{McMillan1986, Kochanek1992, Mardling1995}. Preserving the same $r_{\rm peri,in}$ versus the same $J_{\rm in}$ leads to slightly different tidal evolution in $a_{\rm in}$ and $e_{\rm in}$ (see Fig.~6 and Fig.~16 in \citealt{Mardling1995}).  In our particular example, the final separation after tidal circularization is $\approx$\,1.8 times the minimum periastron separation achieved during the initial rapid tidal decay (see Fig.~4). We further discuss the implications of using two different prescriptions for tidal friction in \S3 and \S5.  

\subsubsection{Further Consideration of Precession}

 For some triples, precession limits the maximum eccentricities achieved by the inner binaries before dynamical tides can effectively decay the orbits.  In the Appendix, we provide the necessary equations for calculating the strengths and rates of precession due to tides and rotation-induced oblateness. Pre-MS primaries are larger and spin more rapidly than MS stars, and so precession can potentially be more important during the pre-MS phase.  However, the strengths of tidal and rotational precession scale with $a_{\rm in}^{-8}$ and $a_{\rm in}^{-5}$, respectively (see Appendix and \citealt{Liu2015}). Because the inner binaries in our triples initially have $a_{\rm in,0}$~$\approx$~10\,-\,100~AU, which are significantly wider than inner binaries with $a_{\rm in,0}$~$<$~1~AU considered in previous studies, the precession strengths are many orders of magnitude weaker.

We examine the effects of precession for our baseline population of triples in which the inner binaries first approach $r_{\rm peri,in}$~$<$~5$R_1$ within $\tau$~$<$~5~Myr. We find 71\% of those inner binaries are limited by precession below $r_{\rm peri,lim}$~$\le$~2.5$R_1$.  In the absence of dynamical tides, these inner binaries would have become tidally disrupted before either tidal or rotational precession could have limited their eccentricity.  We also determine that 80\%,  88\%, and 94\% of the inner binaries are limited by precession below $r_{\rm peri,lim}$~$\le$~3$R_1$, $r_{\rm peri,lim}$~$\le$~4$R_1$, and $r_{\rm peri,lim}$~$\le$~5$R_1$, respectively. In our simulations, the majority of the inner binaries decay to short periods via dynamical tides when $r_{\rm peri,in}$~$\approx$~(3\,-\,5)$R_1$.  The predicted close binary fraction would decrease by only $\approx$10\%\,-\,20\% if we were to include precession in our secular equations.  Because the tidal and rotational precession rates decrease dramatically with $a_{\rm in}$, the effects of precession are negligible for our population of triples with initially wide inner binaries.
  
\section{Results for Main Simulations}

\subsection{Baseline Model A1}

For our baseline model A1 of $N_{\rm trip}$~=~20,000 triples, we numerically integrate the octupole-level secular equations, tidal friction equations, and prescriptions for stellar evolution as described above.  We evolve each system for 200$\tau_{\rm KL}$ in the cases where the inner binaries never approach $P_{\rm in}$~$<$~100~days. In the cases where the inner binaries are tidally captured toward $P_{\rm in}$~$<$~100~days via dynamical tides, we simulate until the inner binaries tidally decay to $e_{\rm in}$~=~0.005 or for 5 Gyr, whichever occurs first. After $\tau$~=~5~Myr, 1251 triples have $P_{\rm in}$~$<$~100~days, 551 of which have $P_{\rm in}$~$<$~10~days ($F_{\rm close}$ = 551$\times$$F_{\rm trip}$/$N_{\rm trip}$ = 0.004). Similarly, after $\tau$~=~5~Gyr, 1964 triples have $P_{\rm in}$~$<$~100~days, 1127 of which have $P_{\rm in}$~$<$~10~days ($F_{\rm close}$ = 0.008).   Only 34 inner binaries achieved very short periastron separations $r_{\rm peri,in}$~$<$~2.5$R_1$ and were subsequently tidally disrupted.  While many inner binaries would have reached $r_{\rm peri,in}$~$<$~2.5$R_1$ in the absence of tidal effects, very efficient dynamical tides in our baseline model prevent a signifiant majority of our systems from crossing this tidal disruption threshold, and so the tidal disruption rate is extremely small. In Fig.~5, we display the period distributions of the inner binaries that evolved toward $P_{\rm in}$~$<$~100~days as well as their corresponding initial inner binary and outer tertiary period distributions. The outer tertiaries in our models are not affected by tides, and their orbital periods do not evolve during their KL oscillations.  We therefore use the terms $P_{\rm out,0}$ and $P_{\rm out}$ interchangeably.

We find that only $F_{\rm close}$~=~0.4\% of systems ($\approx$3\% of triples) contain inner binaries with $P_{\rm in}$~$<$~10~days that formed via KL oscillations and tidal friction during the pre-MS phase.  This is a factor of five times smaller than the observed fraction $F_{\rm close}$~=~2.1\%, indicating the majority of close binaries with outer tertiaries cannot derive exclusively from secular evolution in dynamically stable triples (see more below).  Our simulated fraction $F_{\rm close}$~=~0.4\% is also half of our simple estimate $F_{\rm close}$~=~0.8\% obtained in \S2.2 and in Fig.~2.  This estimate was based on the fraction of systems containing inner binaries that achieved $r_{\rm peri, in}$~$<$~5$R_1$ within $\tau$~$<$~5~Myr.  The discrepancy between our simple estimate and full simulation is due to the additional time lag associated with tidal decay and circularization.  For example, 1251$\times$$F_{\rm trip}$/$N_{\rm trip}$ = 0.9\% of systems contain inner binaries that were tidally captured via dynamical oscillations into $P_{\rm in}$~$<$~100~days by $\tau$~$<$~5~Myr.  This is consistent with the results based on our crude criterion $r_{\rm peri,in}$~$<$~5$R_1$.  However, subsequent energy dissipation via weak-friction equilibrium tides is a substantially slower process, and so only a fraction of those binaries with $P_{\rm in}$~$<$~100~days also happen to tidally decay to $P_{\rm in}$~$<$~10~days by $\tau$~$<$~5~Myr.  The other binaries either remain at $P_{\rm in}$~=~10\,-\,100~days, or tidally decay toward shorter periods $P_{\rm in}$~$<$~10~days during the MS.  In fact, in our baseline model, $\approx$40\% of close binaries that migrate to $P_{\rm in}$~$<$~10~days on the MS were originally tidally captured into eccentric orbits with $P_{\rm in}$~=~10\,-\,30~days during the pre-MS phase (see more below).  This partially explains the slight pile-up of inner binaries with $P_{\rm in}$~=~4\,-\,10~days on the MS relative to the pre-MS (see Fig.~5).

\begin{figure}[t!]
\centerline{
\includegraphics[trim=0.7cm 0.4cm 1.0cm 0.25cm, clip=true, width=3.6in]{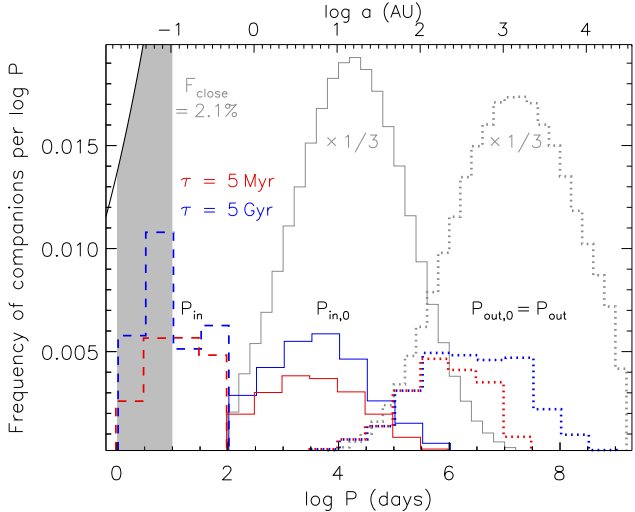}}
\caption{Similar to Fig.~1, but where we have reduced the initial period distributions of inner binaries (thin solid grey) and outer tertiaries (dotted grey) in all triples by a factor of three for display purposes.  After simulating for $\tau$~=~5~Myr (red) and $\tau$~=~5~Gyr (blue), we show the distributions of inner binaries with $P_{\rm in}$~$<$~100~days (dashed) that formed via KL oscillations coupled with tidal energy dissipation. We also display their corresponding initial periods $P_{\rm in,0}$ (solid red/blue) and the periods $P_{\rm out}$ of their outer tertiaries (dotted red/blue).  Compared to the mean value $\langle$log\,$P_{\rm in,0}$\,(days)$\rangle$=~4.3 of the parent distribution,  inner binaries with shorter initial periods log\,$P_{\rm in,0}$\,(days)=~2\,-\,4 produce close binaries more efficiently. Triples with outer tertiaries log\,$P_{\rm out}$\,(days)~$<$~6 form close binaries almost exclusively during the pre-MS phase when $\tau$~$<$~5~Myr.  Meanwhile, triples with wider tertiaries (log\,$P_{\rm out}$~$>$~7.2) require at least $\tau$~$>$~5~Myr for the inner binaries to first approach very large eccentricities in their KL cycles.  By integrating the dashed curves across $P_{\rm close}$~$\equiv$~1~-~10~days, we find $F_{\rm close}$~=~0.4\% and $F_{\rm close}$~=~0.8\% of systems have close binaries that formed via KL cycles and tidal friction by ages $\tau$~=~5~Myr and $\tau$~=~5~Gyr, respectively.  These values are smaller than the observed fraction $F_{\rm close}$~=~2.1\% (filled grey), demonstrating the majority of close binaries with outer tertiaries derives from a different formation channel.}
\end{figure}

We next investigate the initial inner binary period $P_{\rm in,0}$ distributions of systems that produce binaries with $P_{\rm in}$~$<$~100~days (solid red and blue lines in Fig.~5).  Compared to the total population of all inner binaries, which peaks at log\,$P_{\rm in,0}$\,(days)~$\approx$~4.3, the initial periods of systems that produce close binaries are skewed toward shorter periods log\,$P_{\rm in,0}$~$\approx$~2\,-\,4.  This is because inner binaries with initially smaller separations can more readily achieve the necessary $r_{\rm peri,in}$ = $a_{\rm in,0}$(1$-e_{\rm in,max}$) $\lesssim$ 5$R_1$ for tidal friction to be important. For triples with $P_{\rm in,0}$~=~100\,-\,1,000~days, $\approx$11\% form binaries with $P_{\rm in}$~$<$~100~days by $\tau$~=~5~Myr (Fig.~5; note that grey histograms are reduced by a factor of three for display purposes).  This is a factor of 11\%/3\%~$\approx$~4 times larger than that produced by the triple star population as a whole.  If the majority of triples instead had initial inner binaries with $P_{\rm in,0}$~=~100\,-\,1,000~days ($a_{\rm in,0}$~$\approx$~0.5\,-\,2 AU), then the close binary fraction after $\tau$~=~5~Myr would be $F_{\rm close}$~$\approx$~1.7\%, which is consistent with the observations.  Similarly, both \citet{Fabrycky2007} and \citet{Naoz2014} reproduce the close binary fraction via KL cycles and tidal friction only if a significant majority of binaries with $P_{\rm in}$~$<$~10~days derived from only slightly longer initial orbital periods $P_{\rm in,0}$~=~10\,-\,1,000~days.  

Primordial binary formation models do not produce such close companions \citep{Bate1995,Tohline2002,Kratter2011}. A seemingly straightforward explanation is that binaries that fragment within the primordial disks on scales of 10s of AU migrate to $a_{\rm in}$~$\approx$~1~AU due to hydrodynamical forces within the disk, possibly with the assistance of dynamical perturbations from outer tertiaries \citep{Artymowicz1994,Bate1995,Bate2002,Munoz2015}. Yet the efficacy and direction of circumbinary disk migration is highly uncertain \citep{Syer1995,Kratter2010,Satsuka2017}.  Subsequent KL oscillations containing inner binaries that already migrated significantly inward in the disk produce close binaries more efficiently than in our model described entirely by secular evolution.  In \S5, we further discuss energy dissipation in the disk as a mechanism for initially hardening inner binaries in triples. In any case, the processes of triple-star secular evolution coupled with tidal friction, without extra energy dissipation in the disk, cannot reproduce the observed close binary population from the expected initial fragmentation distributions.

The periods of the outer tertiaries in triples that produce close binaries are also weighted toward smaller values compared to their corresponding parent distribution.  In this case, the differences stem primarily from $\tau_{\rm KL}$ (see Eqn.~2).  The majority of triples with log\,$P_{\rm out}$\,(days)~=~4\,-\,6 have rapid KL timescales, and so their inner binaries first reach $r_{\rm peri,in}$~$\lesssim$~5$R_1$ and tidally evolve toward $P_{\rm in}$~$<$~100~days during the pre-MS phase. A negligible fraction of triples with log\,$P_{\rm out}$\,(days)~=~4\,-\,6 produce binaries with $P_{\rm in}$~$<$~100~days after $\tau$~$>$~5~Myr, which is why the dotted red and blue lines in Fig.~5 are nearly identical across this interval.  Meanwhile, triples with log\,$P_{\rm out}$~=~7.0\,-\,8.5 have longer KL timescales and form close binaries predominantly on the MS.  The widest tertiaries with log\,$P_{\rm out}$~$>$~8.6 have KL timescales too long to produce any close binaries, even during our 5 Gyr simulation. 

While common-proper-motion companions to solar-type field MS stars are observed up to log\,$P$~=~9.3 \citep{Raghavan2010}, tertiary companions to close MS binaries with $P_{\rm in}$~$<$~10~days span only up to log\,$P_{\rm out}$~$<$~8.7 \citep{Tokovinin2008}.  The latter is consistent with our models.  The absence of extremely wide companions to close binaries indicates that slightly closer tertiaries with log\,$P_{\rm out}$~$<$~8.7 are required to dynamically harden the inner binaries within their KL timescales.  Unfortunately, the observed difference in the maximum period of MS binaries (log\,$P$~$=$~9.3) and of tertiary companions to close MS binaries (log\,$P_{\rm out}$~$=$~8.7) is small and may instead be due to the limited sample sizes and/or systematic uncertainties in de-projecting the observed binary component offsets into true orbital separations. Nevertheless, we expect the difference in the maximum orbital periods to be even larger and more apparent for pre-MS systems.  In particular, we predict that T~Tauri stars that have close companions should exhibit an absence of extremely wide tertiaries beyond log\,$P_{\rm out}$~$>$~7.2 (Fig.~5).  The observed frequency of wide companions with log\,$P$~=~7.0\,-\,8.5 to T~Tauri stars is already a factor of $\approx$2\,-\,3 times larger than the frequency of such wide companions to solar-type MS stars in the field \citep{Duchene2007, Connelley2008, Tobin2016a}.  The predicted deficit of extremely wide tertiaries to T~Tauri close binaries should be even more obvious.  With a large enough sample, future observations should soon be able to confirm or reject our hypothesis that T~Tauri close binaries exhibit a deficit of wide tertiaries beyond log\,$P_{\rm out}$\,(days)~$\gtrsim$~7.0 ($a_{\rm out}$~$\gtrsim$~1,000 AU). Although we conclude that close binaries cannot derive exclusively from secular evolution, we suggest the subset of close binaries with inclined, wide outer tertiaries $a_{\rm out}$~$\gtrsim$~200~AU may in fact derive from KL oscillations (see \S5).  We therefore still expect the predicted trend whereby the maximum orbital period of outer tertiaries to close binaries increases with age.  If this effect is observed, it simply suggests that close binaries with wide outer tertiaries $a_{\rm out}$~$\gtrsim$~200~AU derive from KL cycles, not that most close binaries derive from KL cycles.

In Fig.~6, we show the eccentricities and periods of the inner binaries that tidally decayed to $P_{\rm in}$~$<$~100~days by ages $\tau$~=~5~Myr (top panel) and $\tau$~=~5~Gyr (bottom panel).  We also display lines of constant orbital angular momentum $J_{\rm in}$.  Inner binaries tidally decay along fixed $J_{\rm in}$ in the weak-friction equilibrium tide model assuming $J_{\rm in}$~$\gg$~$J_1$.  As indicated above, we simulate 214 pre-MS inner binaries with $P_{\rm in}$~$>$~10~days that lie above the middle red $J_{\rm in}$ curve in the top panel of Fig.~6. These eccentric binaries with $e_{\rm in}$~$\approx$~0.6\,-\,0.8 and $P_{\rm in}$~=~10\,-\,30~days tidally evolve to $e_{\rm in}$~=~0 and $P_{\rm in}$~$\approx$~4\,-\,10~days by $\tau$~=~5~Gyr, contributing 214$\times$$F_{\rm trip}$/$N_{\rm trip}$ $\approx$ 0.0016 to the close binary fraction $F_{\rm close}$.  The simulated close binary fraction increases by $\delta$$F_{\rm close}$ = 0.004 between $\tau$~=~5~Myr and $\tau$~=~5~Gyr.  We therefore find that 0.0016/0.004 = 40\% of this growth is due to inner binaries initially captured by $\tau$~$<$~5~Myr into eccentric orbits slightly beyond $P_{\rm in}$~$>$~10~days that subsequently decay toward shorter periods $P_{\rm in}$~$<$~10~days on the MS.

\begin{figure}[t!]
\centerline{
\includegraphics[trim=3.2cm 0.3cm 4.2cm 0.2cm, clip=true, width=3.55in]{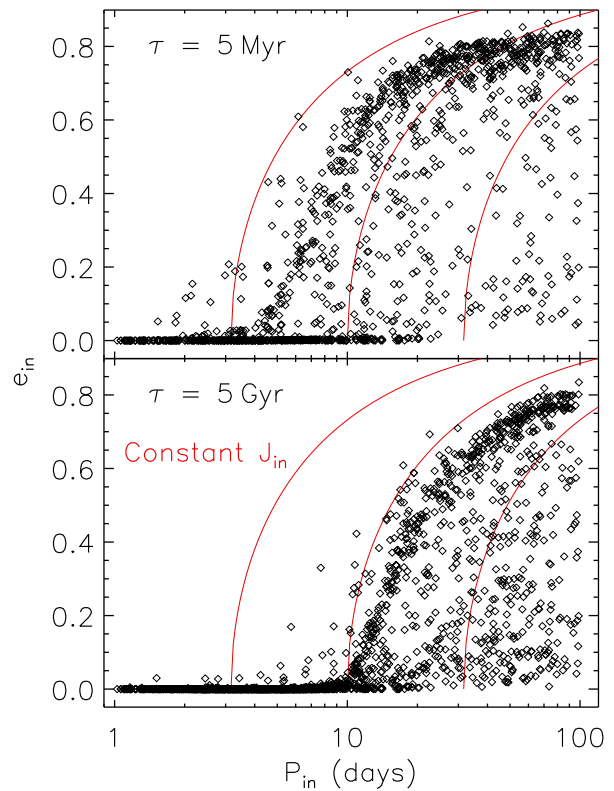}}
\caption{Eccentricities $e_{\rm in}$ vs. periods $P_{\rm in}$ of the inner binaries that tidally decayed to $P_{\rm in}$~$<$~100~days in our baseline model by ages $\tau$~=~5~Myr (top) and $\tau$~=~5~Gyr (bottom). The red curves indicate lines of constant orbital angular momentum $J_{\rm in}$.  By setting the scale factor to $F_{\rm tide}$~=~20 in the weak-friction equilibrium-tide model, we reproduce the observed circularization periods of $P_{\rm circ}$~$\approx$~6~days and $P_{\rm circ}$~$\approx$~11~days for the pre-MS and field MS binary populations, respectively.   There is a concentration of $e_{\rm in}$~=~0 binaries across $P_{\rm circ}$~$<$~$P_{\rm in}$~$<$~20~days due to tidal circularization when the pre-MS primary was $R_1$~=~4\,-\,10\,\Rsun.  Observations also reveal circularized binaries beyond $P_{\rm in}$~$>$~$P_{\rm circ}$, which further demonstrates that a non-negligible fraction of close binaries originally migrated during the early pre-MS phase.  We find a few eccentric binaries across $P$~$<$~$P_{\rm circ}$, which were not excited and sustained toward larger eccentricities by their outer tertiaries as suggested by previous studies, but instead were only recently tidally captured into short-period eccentric orbits immediately prior to the snapshots at $\tau$~=~5~Myr and $\tau$~=~5~Gyr shown here.  Inner binaries with $P_{\rm in}$~$>$~$P_{\rm circ}$ are concentrated in a locus that asymptotes toward our adopted transition eccentricity $e_{\rm trans}$~=~0.8 between the dynamical oscillation tidal regime and the weak-friction tidal regime.  Inner binaries in triples with $P_{\rm circ}$~$<$~$P_{\rm in}$~$<$~100~days are weighted toward large eccentricities, which is consistent with observations.}
\end{figure}

There are four other noteworthy features of Fig.~6:
\begin{itemize}
\item{After $\tau$~=~5~Myr, the majority of pre-MS binaries with $P_{\rm circ}$~$<$~5~days have been tidally circularized while a substantial fraction of binaries with $P_{\rm circ}$~$>$~5~days remain eccentric.  This is consistent with the observed circularization period $P_{\rm circ}$~$\approx$~6~days of close pre-MS binaries in young star-forming environments \citep{Meibom2005}.   Similarly, the simulated population after $\tau$~=~5~Gyr exhibits a circularization period of $P_{\rm circ}$ = 11 days.  This matches the observed circularization period of solar-type MS binaries in the field \citep{Meibom2005,Raghavan2010}.  Only by setting $F_{\rm tide}$~=~20 in our baseline model for weak-friction equilibrium tides can we reproduce the observed circularization periods in both the pre-MS and MS populations.}

\item{We find a locus of binaries extending from $e_{\rm in}$~=~0 at $P_{\rm circ}$ to $e_{\rm in}$~=~0.8 at $P_{\rm in}$~=~100~days. The loci in both the pre-MS and MS populations closely resembles the curves of constant orbital angular momentum $J_{\rm in}$.  Inner binaries that approach small periastron separations in their KL cycles undergo rapid tidal decay via non-radial dynamical oscillations until the eccentricity reaches $e_{\rm trans}$~=~0.8.  Subsequent evolution via weak-friction equilibrium tides is a slower process, especially for wider binaries.  This is the reason for the large concentration of binaries near $e_{\rm trans}$~$=$~0.8 at long orbital periods.  Setting  $e_{\rm trans}$ to a smaller value would cause the loci to asymptote at smaller eccentricities (see \S3.2).  In reality, both weak-friction equilibrium tides and dynamical oscillations contribute to tidal friction across all eccentricities.  Our model simply encapsulates the notion that the latter process dominates at larger eccentricities. We expect binaries across $P_{\rm circ}$~$<$~$P_{\rm in}$~$<$~100~days that originally dynamically evolved due to KL cycles and tidal friction to have systematically larger eccentricities than binaries without tertiary companions.  \citet{Raghavan2010} indeed confirm that binaries with $P_{\rm circ}$~$<$~$P_{\rm in}$~$<$~100~days that trace the upper envelope of the period-eccentricity relation have tertiary companions while the majority of systems with smaller eccentricities are solitary binaries (see their Fig.~14).  Heartbeat stars, which exhibit small-amplitude photometric variations due to dynamical oscillations, have also been demonstrated to be binaries that trace the upper envelope of the eccentricity-period relation \citep{Shporer2016}.  In their sample,  \citet{Shporer2016} find that heartbeat stars extend from $e$~$\approx$~0.3 at $P$~$\approx$~10~days to $e$~$\approx$~0.9 at $P$~$\approx$~100~days (see their Fig.~6).  Due to their concentration along the upper envelope of the period-eccentricity relation, we expect the majority of heartbeat binaries to harbor outer tertiary companions, many of which likely migrated via KL oscillations and tidal friction.  The fact that heartbeat stars are observed in binaries with eccentricities down to $e$~=~0.3 further suggests that dynamical oscillations at least partially contributes to tidal friction across intermediate eccentricities $e$~=~0.3\,-\,0.8.}

\item{There is a small population of circularized binaries beyond $P_{\rm in}$~$>$~$P_{\rm circ}$, extending up to $P_{\rm in}$~$\approx$~20~days ($a_{\rm in}$~$\approx$~35\,\Rsun). These inner binaries were initially tidally captured into short orbits via dynamical oscillations during the very early pre-MS phase when $R_1$~=~4\,-\,10\,\Rsun\  ($\tau$~$\lesssim$~0.2~Myr).  Their orbits quickly circularized via weak-friction equilibrium tides at systematically larger separations due to the larger sizes of their young pre-MS primaries.  Very large samples of spectroscopic \citep{Halbwachs2003,Meibom2005} and eclipsing \citep{Khaliullin2010,Moe2015,Kirk2016} MS binaries also reveal a small population of circularized binaries beyond $P_{\rm circ}$, albeit not as numerous as in our simulation.   Our model A8, which incorporates a smaller pre-MS primary $R_1$~$<$~3\,\Rsun, yields a population of pre-MS circularized binaries that is more consistent with observations (see \S3.2)}. The observed population of circularized binaries beyond $P_{\rm circ}$ demonstrates that at least some close binaries migrated inward during the early pre-MS phase.

\item{We find several slightly eccentric $e_{\rm in}$~$\approx$~0.01\,-\,0.3 binaries with  periods $P_{\rm in}$~$<$~$P_{\rm circ}$ below the circularization period in both the $\tau$~=~5~Myr pre-MS and $\tau$~=~5~Gyr MS populations.  Spectroscopic surveys also discover slightly eccentric binaries shortward of the circularization period, nearly all of which have detected tertiary companions \citep{Meibom2005,Raghavan2010}.  However, previous studies have suggested that KL oscillations excite and sustain the eccentricities, counteracting tidal effects \citep{Meibom2005,Raghavan2010,Anderson2017}.  For instance, \citet{Anderson2017} modeled a particular eccentric binary with $P_{\rm in}$~=~15~days under the assumption that secular evolution via KL cycles with a tertiary counteracts tidal and general relativistic precession.  In their model, KL oscillations pump the eccentricity and obliquity of the inner $P_{\rm in}$~=~15~day binary to its currently observed non-zero values.  While KL oscillations may affect the eccentricity and obliquity evolution of binaries beyond $P_{\rm in}$~$\gtrsim$~12~days, as in the case study investigated by \citet{Anderson2017}, we expect tidal friction to dominate at shorter periods $P_{\rm in}$~$<$~$P_{\rm circ}$~$\approx$~6\,-\,11~days.  The reason for slightly eccentric binaries with $P$~$<$~$P_{\rm circ}$ in our simulations is because they were only recently tidally captured into short-period orbits.  Their ``tidal age'', i.e. the time they have spent at short periods experiencing tidal friction, is therefore a small fraction of their true age.  If a larger fraction of close binaries were initially tidally captured into short orbits while on the MS, then we would expect a larger fraction of close binaries with $P_{\rm in}$ $<$ $P_{\rm circ}$ to exhibit non-zero eccentricities.  Observations reveal only a handful of short-period eccentric binaries, consistent with our simulations and further indication that the majority of close binaries originally migrated to short periods during the pre-MS phase.}
\end{itemize}

\begin{figure}[t!]
\centerline{
\includegraphics[trim=0.8cm 0.2cm 1.2cm 0.1cm, clip=true, width=3.45in]{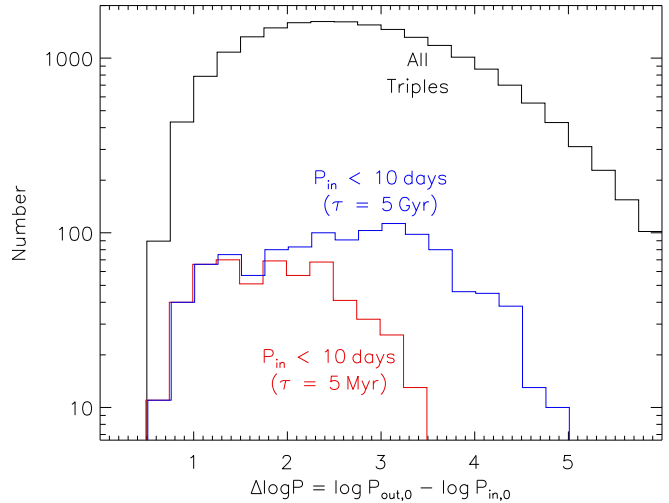}}
\caption{The distribution of initial triple-star hierarchies $\Delta$log$P$~=~log($P_{\rm out,0}$/$P_{\rm in,0}$) in our baseline model (black) and the subset that produce close inner binaries with $P_{\rm in}$~$<$~10~days by ages $\tau$~=~5~Myr (red) and $\tau$~=~5~Gyr (blue).  While only $\approx$3\% of triples in our baseline model form close binaries via KL oscillations and tidal friction during the pre-MS phase, triples in initially marginally stable orbits $\Delta$log$P$ = 0.6\,-\,1.5 produce close binaries more efficiently due to octupole-level effects.}
\end{figure}

In Fig.~7, we show the distribution of initial triple-star hierarchies $\Delta$log$P$~=~log\,$P_{\rm out,0}-$log\,$P_{\rm in,0}$ in our baseline model.  We also display the subset of triples that produce binaries with $P_{\rm in}$~$<$~10~days by ages $\tau$~=~5~Myr and $\tau$~=~5~Gyr. Triples with highly hierarchical configurations, i.e., larger $\Delta$log$P$, have systematically longer KL timescales (Eqn.~2).  Triples with $\Delta$log$P$~$>$~3.5 cannot form close binaries during the pre-MS phase, and triples with $\Delta$log$P$~$>$~5.0 cannot form close binaries even during our $\tau$~=~5~Gyr simulation. By $\tau$~=~5~Gyr, 1127/20000~=~5.6\% of triples in our baseline model form binaries with $P_{\rm in}$~$<$~10~days. Meanwhile, $\approx$9\% of triples in initially marginally stable configurations $\Delta$log$P$ = 0.6\,-\,1.5 produce close binaries, nearly all of which migrate to $P_{\rm in}$~$<$~10~days by $\tau$~=~5~Myr.  Triples in marginally stable configurations, which are more heavily affected by octupole-level effects, produce close binaries at nearly twice the rate of the triples in our overall population.  Octupole-level effects widen the parameter space of mutual inclinations that can produce close binaries, but only mildly.  The enhancement is at most a factor of two for triples with $\Delta$log$P$~$<$~1.5 ($a_{\rm out,0}$/$a_{\rm in,0}$~$<$~10), and only $\approx$12\% of the stable triples in our baseline model have such configurations (see also \S2.1). For close binaries that formed through KL oscillations and tidal friction, we expect the majority to have originated in triples that had intermediate hierarchies $\Delta$log$P$~=~1.5\,-\,3.5 ($a_{\rm out,0}$/$a_{\rm in,0}$ = 10\,-\,200).

\subsection{Ancillary Models A2-A11}

We next compare our ancillary models A2-A11 to the baseline model A1.  The model parameters are outlined in \S2, but to summarize: models A2-A5 incorporate different initial separation distributions, models A6-A7 assume different initial eccentricities distributions,  model A8 sets the maximum radius of the pre-MS primary to $R_1$~=~3\,\Rsun, and models A9-A11 utilize different prescriptions for tidal energy dissipation.  In Table 1, we show the results of the models where the four columns correspond to (1) the model designation, (2) how the value of the variable parameter changed from the baseline model to the listed model, (3) the simulated close binary fraction $F_{\rm close}$ after $\tau$~=~5 Myr, and (4) $F_{\rm close}$ after $\tau$~=~5 Gyr.

{\bf Models~A2-A5:} Changes in the initial separation distributions most dramatically affect the simulated close binary fractions.  The directions and magnitudes of the changes reflect what we would expect based on Fig.~5 and discussion in \S3.1.  Triples with outer tertiaries that are weighted toward smaller separations $a_{\rm out,0}$ have shorter KL timescales, and so can produce more close binaries via KL oscillations and tidal friction during the finite pre-MS and MS phases.  Similarly, triples with inner binaries distributed toward smaller separations $a_{\rm in,0}$ can more readily achieve sufficiently small periastron separations $r_{\rm peri,in}$~$<$~5\,$R_1$ for tidal energy dissipation to be efficient.  Shifting the peak in the separation distribution of the inner binaries by a factor of three changes the simulated close binary fractions during both the pre-MS and MS phases by a factor of $\approx$\,1.6 (see Models A2 and A3 in Table~1).  A large fraction of simulated close binaries in all of our models derive from triples with $a_{\rm in,0}$~=~1\,-\,10 AU that span the short-period tail of the parent distribution (see Fig.~5).  Disk fragmentation likely occurs on slightly larger scales $a_{\rm in,0}$~=~50\,-\,100 AU \citep{Bate1995,Rafikov2005,Kratter2008}, and so inner binaries with $a_{\rm in,0}$~=~1\,-\,10~AU most likely have already migrated slightly inward due to hydrodynamical forces in the disk (see \S3.1 and \S5). Systematically shifting the inner binaries towards smaller separations increases $F_{\rm close}$, and so inward disk migration of the inner binaries prior to KL oscillations with an outer tertiary may assist in the formation of close binaries (see further discussion in \S5).

\begin{figure}[t!]\footnotesize
{\bf Table 1:} Results for Models A1\,-\,A11  \\
\vspace*{-0.55cm}
\begin{center}
\begin{tabular}{|c|l|c|c|}
\hline
  &    &  $F_{\rm close}$ & $F_{\rm close}$ \\
  Model &  Parameter Change from A1 & ($\tau$\,=\,5\,Myr) & ($\tau$\,=\,5\,Gyr) \\
\hline
 A1     &    -       & 0.4\%  & 0.8\% \\
\hline 
 A2      & $\mu_{\rm loga; in}$~=~1.5~$\rightarrow$~1.0 & 0.6\% & 1.1\% \\
\hline
 A3      & $\mu_{\rm loga; in}$~=~1.5~$\rightarrow$~2.0 & 0.2\% & 0.5\% \\
\hline
 A4      & $\mu_{\rm loga; out}$~=~2.8~$\rightarrow$~2.3 & 0.5\% & 1.0\% \\
\hline
 A5      & $\mu_{\rm loga; out}$~=~2.8~$\rightarrow$~3.3 & 0.3\% & 0.6\% \\
\hline
 A6      & $e_{\rm in}$~=~0.01~$\rightarrow$~$p$($e_{\rm in}$)~$\propto$~$e_{\rm in}^{-0.8}$ & 0.3\% & 0.6\% \\
\hline
 A7      & $p$($e_{\rm out})$~=~1~$\rightarrow$~2$e_{\rm out}$ & 0.5\% & 0.9\% \\
\hline
 A8      & max($R_1$)~=~10\Rsun~$\rightarrow$~3\Rsun~ & 0.6\% & 0.9\% \\
\hline
 A9      & $e_{\rm trans}$~=~0.8~$\rightarrow$~0.5  & 0.6\% & 0.9\% \\
\hline
 A10      & $T_2$~=~0.3$\eta^{-2}$~$\rightarrow$~0.03$\eta^{-2}$  & 0.2\% & 0.6\% \\
\hline
 A11     & $F_{\rm tide}$~=~20~$\rightarrow$~1  & 0.2\% & 0.7\% \\
\hline
\end{tabular}
\end{center}
\end{figure}

{\bf Models~A6-A7:} Increasing the initial eccentricities of the inner binaries slightly decreases the close binary fraction.  In our baseline model in which all inner binaries initially have nearly circular orbits ($e_{\rm in,0}$~=~0.01), the initial arguments of periastron $\omega_{\rm in,0}$ and $\omega_{\rm in,0}$ have a negligible effect on the evolution of the system.  For the small fraction of triples with $e_{\rm in,0}$~$\gtrsim$~0.3 in our Model A6, however, only a certain combination of $\omega_{\rm in,0}$ and $\omega_{\rm out,0}$ can lead to KL oscillations that approach very large eccentricities $e_{\rm in}$~$\gtrsim$~0.99 (see discussion of this effect in \citealt{Katz2011}).  By changing the eccentricity distribution of the outer tertiaries from a uniform distribution to a thermal eccentricity distribution, the close binary fraction slightly increases (Model A7).  Weighting the outer eccentricities $e_{\rm out}$ toward larger values systematically decreases the KL timescales (Eqn.~2) and increases the strength of the octupole level (Eqn.~6).  

{\bf Model~A8:} Setting the maximum radius of the pre-MS primary to $R_1$~=~3\,\Rsun\ has two main effects.  First, the population of pre-MS circularized binaries beyond $P_{\rm in}$~$>$~$P_{\rm circ}$~$\approx$~6~days shown in Fig.~6 is considerably diminished, but nonetheless still existent.  In our baseline Model A1, we simulate pre-MS circularized binaries out to $P_{\rm in}$~$\approx$~20~days, but in our Model~A8, circularized pre-MS binaries extend only to $P_{\rm in}$~$\approx$~10~days.  The latter is more consistent with observations of close binaries \citep{Halbwachs2003,Meibom2005,Kirk2016}, indicating either pre-MS stars accrete most of their mass while $R_1$~$<$~3\,\Rsun\ \citep{Larson1969,Hosokawa2009} and/or secular evolution during the first $\tau$~$\approx$~0.3~Myr is suppressed due to interactions with the massive natal disks (see more below).  Second, adopting a smaller pre-MS primary increases our simulated close binary fraction, especially  during the pre-MS phase.  In our baseline Model~A1, inner binaries that tidally decay to short periods during the first $\tau$~$<$~0.3~Myr when the pre-MS primary is large ($R_1$~$>$~3\,\Rsun) settle into orbits with $P_{\rm in}$~=~10\,-\,100 days. Such large pre-MS primaries cannot accommodate short-period companions with $P_{\rm close}$~$<$~10~days.  By reducing the maximum size of very young pre-MS primaries, many inner binaries that originally migrated to $P_{\rm in}$~=~10\,-\,100~days now reach $P_{\rm in}$~$<$~10~days.

{\bf Model~A9:} Decreasing the transition eccentricity between the two tidal mechanisms from $e_{\rm trans}$ = 0.8 to $e_{\rm trans}$ = 0.5 moderately increases $F_{\rm close}$, especially during the pre-MS phase.  As discussed in \S2.3, our treatment of tidal friction via dynamical oscillations preserves $r_{\rm peri,in}$.  Subsequent tidal energy dissipation via weak-friction equilibrium tides preserves the orbital angular momentum, and so $r_{\rm peri,in}$ increases slightly. By decreasing $e_{\rm trans}$, the final separations $a_{\rm in}$ of the inner binaries are reduced.  Moreover, tidal friction via dynamical oscillations operates on significantly faster timescales than weak-friction equilibrium tides.  Reducing $e_{\rm trans}$ does not significantly affect the number of inner binaries that achieve $P_{\rm in}$~$<$~100~days, but instead simply shifts a larger fraction of inner binaries with $P_{\rm in}$~$<$~100~days toward even shorter periods $P_{\rm in}$~$<$~10~days during the pre-MS phase. In our baseline model A1, a large fraction of inner binaries circularize on the MS across $P_{\rm in}$~=~4\,-\,10 days (see enhancement in Fig.~4).  Decreasing the transition eccentricity to $e_{\rm trans}$~=~0.5 shifts the enhancement to $P_{\rm in}$~=~2\,-\,6 days. Close binaries with $P$~=~2\,-\,6~days are more likely to have outer tertiaries than binaries with $P$~=~6\,-\,10~days \citep{Tokovinin2006}.  Our Model A9 is more consistent with this observed trend, further suggesting dynamical tides are important even at moderate eccentricities $e_{\rm in}$~=~0.5\,-\,0.8.

{\bf Model~A10:} Reducing the efficiency of dynamical tides by an order of magnitude from $T_2$~=~0.3$\eta^{-2}$ to $T_2$~=~0.03$\eta^{-2}$, as motivated by more recent calculations of energy dissipation via dynamical oscillations \citep{McMillan1986,Mardling1995,Lai1997}, decreases the simulated close binary fraction. In our baseline Model~A1, a negligible fraction of inner binaries are tidally disrupted because dynamical tides quickly dissipate enough energy to prevent them from crossing $r_{\rm peri,in}$~$\lesssim$~2.5$R_1$.  In our Model A10 with weaker dynamical tides, however, $\approx$15\% of inner binaries that achieve $r_{\rm peri,in}$~$<$~5$R_1$ reach $r_{\rm peri,in}$~$<$~2.5$R_1$ and are subsequently tidally disrupted.  Inner binaries that achieved $r_{\rm peri,in}$~$\approx$~(4\,-\,5)$R_1$ in our baseline model now evolve slightly differently with weaker dynamical tides.  For instance, some inner binaries that decayed to $P_{\rm in}$~$<$~10~days no longer become close binaries in our Model~A10.  Conversely, some inner binaries that tidally decayed to $P_{\rm in}$~=~10\,-\,30~days in our baseline model now reach slightly smaller periastron separations $r_{\rm peri,in}$~$\approx$~(3\,-\,4)$R_1$ and subsequently decay to $P_{\rm in}$~$<$~10~days.  By combining these various effects, the net result of considering weaker, more realistic dynamical tides is that the simulated pre-MS close binary fraction decreases by $\approx$40\%.

{\bf Model~A11:} For our final model, decreasing the scaling factor for weak-friction equilibrium tides to the nominal value $F_{\rm tide}$~=~1 reduces the simulated close binary fraction, especially during the pre-MS phase.  Many of the inner binaries that tidally decayed toward smaller eccentricities and $P_{\rm in}$~$<$~10~days by $\tau$~=~5~Myr in our baseline model are now left at slightly longer periods $P_{\rm in}$~=~10\,-\,30~days in our Model~A11.  However, by setting $F_{\rm tide}$~=~1, the simulated circularization periods are reduced to $P_{\rm circ}$~=~3~days and $P_{\rm circ}$~=~6~days for population ages of $\tau$~=~5~Myr and $\tau$~=~5~Gyr, respectively. These results are measurably discrepant with the observed circularization periods of $P_{\rm circ}$~=~6~days and $P_{\rm circ}$~=~11~days for pre-MS binaries and solar-type MS binaries in the field, respectively \citep{Meibom2005,Raghavan2010}.  As noted in \citet{Meibom2005}, the discrepancies between the observed and simulated circularization periods further demonstrate that tidal friction in solar-type binaries is not fully understood. Nevertheless, the uncertainties in tidal friction alone cannot account for the differences between the simulated close binary fractions and the observed value $F_{\rm close}$~=~2.1\%.  

Considering all the sources of uncertainty incorporated into our models and discussed above, we estimate that only 0.4\%\,$\pm$\,0.2\% of systems produce close binaries with $P_{\rm in}$~=~1\,-\,10~days exclusively through KL cycles coupled with tidal friction by age $\tau$~=~5~Myr.  Given the observed $F_{\rm close}$~=~2.1\%, then only 20\%\,$\pm$\,10\% of close binaries derive strictly from secular evolution in triples and tidal energy dissipation during the pre-MS phase.  Other physical processes are required to explain the formation and properties of close binaries, which we further discuss in \S4 and \S5.   We also find that only an additional (20\,$\pm$\,10)\% of close binaries are formed exclusively through KL oscillations and tidal friction during the MS phase.  This small increase in the close binary fraction between $\tau$~=~5~Myr and $\tau$~=~5~Gyr is consistent with the observational constraints (see \S1).

There are other sources of uncertainty not directly incorporated into our simulations. First, tidal precession may decrease the predicted number of close pre-MS binaries that derive from KL cycles by $\approx$10\%\,-\,20\% (see \S2.3.5).  Second, in all of our simulations, we assumed $F_{\rm trip}$~=~15\% of systems are born in stable triples in which the inner companion forms via disk fragmentation, the outer tertiary forms via core fragmentation, and the mutual inclinations are isotropically distributed.  In reality, some triples, especially those with $a_{\rm out}$~$\lesssim$~100~AU, have coplanar configurations \citep{Tokovinin2017}, suggesting both companions derive from disk fragmentation.  The period and inclination distributions of pre-MS triples have yet to be accurately measured, but the fraction of young systems in triples with random orientations is likely smaller than $F_{\rm trip}$~=~15\%.  Third, massive natal disks likely suppress secular evolution via KL cycles during the first $\tau$~$\approx$~1~Myr.  While delaying secular evolution until $R_1$~$<$~3$R_1$ ($\tau$~$\approx$~0.3~Myr) mildly increases the predicted close binary fraction (see Model A8), further delay likely decreases the predicted number of close binaries.  Next, we incorporated optimistic secular equations for dynamical tides, while more realistic tidal evolution, which includes chaos in the subsequent periastron passages \citep{Mardling1995}, may decrease the efficiency of energy dissipation (see also Model A10).  Finally, we used a broad Gaussian to describe the initial inner binary distribution. Some of the inner binaries in our models have $a_{\rm in,0}$~$<$~10~AU, and many simulated close binaries derive from this short-period tail (see Fig.~5). Primordial binaries with $a_{\rm in,0}$~$<$~10~AU are likely unphysical, and so our Model A3 with even wider initial inner binaries may be more realistic.  The various effects considered above would tend to reduce the expected number of close pre-MS binaries, which strengthens our overall conclusion that most close binaries cannot derive exclusively from KL cycles and tidal friction without extra energy dissipation in the disk.

\section{Compact Unstable Coplanar Triples}

We next explore an alternative channel for producing close binaries that involves triples initially born in unstable configurations.  We specifically investigate unstable triples where one of the components is dynamically ejected, which leaves behind solitary binaries in eccentric orbits with slightly reduced separations.  There are three main processes by which triples are born in unstable configurations. First, the inner and outer companions form via disk and core fragmentation, respectively, but at the wings of their corresponding separation distributions such that $a_{\rm in,0}$~$\sim$~$a_{\rm out,0}$~$\sim$~100~AU. In \S2.1.2, we discussed how ejection of one of the components in these types of unstable triples leaves behind solitary binaries with $a$~$\sim$~40~AU that follow a thermal eccentricity distribution $p$~=~2$e$ \citep{Valtonen2006}.  In this scenario, a negligible fraction of the remaining binaries achieve small enough periastron separations for tidal energy dissipation to be effective.  Second, both the inner and outer companions derive from core fragmentation on large scales $a_{\rm in,0}$~$\sim$~$a_{\rm out,0}$~$\sim$~600~AU.  Because the unstable triples in this scenario are even wider than in the previous channel, the resulting close binary population is even smaller. Finally, unstable triples can form when two companions fragment in the disk on scales of $a_{\rm in,0}$~$\sim$~$a_{\rm out,0}$~$\sim$~30~AU.  These unstable triples are more compact and have a higher probability of forming close binaries.  Moreover, while we expect triples that derive from core + disk fragmentation or core + core fragmentation to have isotropic orientations, triples in which both companions derive from disk fragmentation should have coplanar orbits.   Indeed, the majority of compact triples with $a_{\rm out}$~$\lesssim$~50~AU  have such coplanar configurations \citep{Sterzik2002,Borkovits2016,Tokovinin2017}.  Dynamical ejection of one of the components in initially unstable coplanar triples produce solitary binaries that are weighted toward even larger eccentricities compared to a thermal eccentricity distribution (\citealt{Valtonen2006}; see also below).  In this section, we simulate the formation of close binaries that derive from the dynamical disruptions of unstable coplanar triples with $a_{\rm in,0}$~$\sim$~$a_{\rm out,0}$~$\sim$~30\,-\,100\,AU.  We do not numerically integrate the dynamical evolution of individual triples, but instead rely on statistical distributions that describe the ensemble population as discussed in \citet{Valtonen2006}.  

Semi-analytic models suggest that cores destined to form stars more massive than $\gtrsim$\,1.0\,\Msun\ may experience a brief period of disk instability leading to fragmentation \citep{Kratter2008}. It is expected that disk fragmentation will typically lead to the formation of stellar or brown dwarf companions \citep{Kratter2010,Stamatellos2009}.  However, a robust estimate for the fraction of disks that undergo fragmentation remains elusive, as does the mass spectrum and number of fragments produced. Even if multiple fragmentation events in the primordial disk are uncommon, e.g., two stellar companions form within the disk in only $F_{\rm TripleDisk}$~$\approx$~10\% of the systems, then our dynamical ejection scenario can still measurably contribute to the observed population of binaries with $P$~$<$~100~days (see below). 

For our first model B1, we select both inner and outer companions from initial probability distributions used to describe the disk population of companions in \S2.1.  We select both companions independently from a log-normal separation distribution with mean of $\mu_{\rm loga}$~=~1.5 (30~AU) and dispersion of $\sigma_{\rm loga}$~=~0.8.  We set the primary mass to $M_1$~=~1~\Msun, and we choose the mass ratios of both companions from a uniform distribution across the interval $q$~=~$M_{\rm comp}$/$M_1$~=~0.1\,-\,1.0.  We select the initial eccentricities of both companions independently from a distribution $p$~$\propto$~$e^{\rm -0.5}$ across $e$~=~0.01\,-\,0.99 weighted toward small values.  We set the mutual inclination to $i_{\rm tot,0}$~=~0 for our triples in which both companions derive from disk fragmentation.  We also consider a second model B2 that has the same initial conditions as described above except for a separation distribution that peaks at a slightly wider value $\mu_{\rm loga}$~=~2.0 (100~AU; similar to disk population in our A3 model).  In both the B1 and B2 models, $\approx$56\% of the generated triples are dynamically unstable according to Eqn.~1.  For our simulations in this section, we consider only the triples that are born in initially unstable configurations and are eventually disrupted.  We also assume that one of the companions is ejected, and so the primary with mass $M_1$~=~1\Msun\ always remains gravitationally bound to the remaining companion. The probability of ejecting the most massive component $M_1$ is small, and we are mainly interested in the properties of companions to $M_1$~=~1\,\Msun\ primaries.

For an unstable coplanar triple, the inner companion with mass $M_{\rm in}$ is only marginally inside the outer companion with mass $M_{\rm out}$.  The probability of ejecting $M_{\rm out}$ (compared to ejecting $M_{\rm in}$) is therefore primarily dependent on mass \citep{Valtonen2006}:

\begin{equation}
 p_{\rm out} = \frac{M_{\rm out}^{-3}}{M_{\rm out}^{-3}+M_{\rm in}^{-3}}.
\end{equation}

\noindent We can write $p_{\rm in}$ simply by switching the variables $M_{\rm out}$ and $M_{\rm in}$ in Eqn.~16.  Using a Monte Carlo technique, we eject the outer companions a fraction $p_{\rm out}$ of the time and the inner companions for the remaining fraction $p_{\rm in}$~=~1\,$-$\,$p_{\rm out}$. 

The total mass of the triple is $M_{\rm tot}$~=~$M_1$\,+\,$M_{\rm in}$\,+\,$M_{\rm out}$ and the initial energy of the system is:

\begin{equation}
 E_{\rm 0} = -\frac{{\rm G} M_{\rm in} M_1}{2 a_{\rm in}} - \frac{{\rm G} M_{\rm out}(M_1 + M_{\rm in})}{2 a_{\rm out}}
\end{equation}

\noindent  If the outer companion is dynamically ejected, it will leave the system with a velocity $v_{\rm out}$.  In the coplanar case, the distribution of velocities peaks at (Eqn.~7.20 in \citealt{Valtonen2006}):

\begin{equation}
 v_{\rm out} = \Big[ \frac{2 (M_{\rm tot} - M_{\rm out})}{5 M_{\rm out} M_{\rm tot}} |E_{\rm 0}| \Big]^{\nicefrac{1}{2}}
\end{equation}

\noindent and the final energy of the system (including the ejected companion) is:

\begin{equation}
 E_{\rm f} = -\frac{{\rm G} M_{\rm in} M_1}{2 a_{\rm f}} + \frac{M_{\rm out} v_{\rm out}^2}{2}.  
\end{equation}

\noindent By setting $E_{\rm f}$~=~$E_{\rm 0}$, we solve for the final separation $a_{\rm f}$ of the remaining solitary binary with component masses $M_1$ and $M_{\rm in}$.  If instead the inner companion was ejected, we switch the variables $M_{\rm out}$ and $M_{\rm in}$ in Eqns.~18\,-\,19 and solve for the velocity $v_{\rm in}$ of the ejected inner companion and the final separation $a_f$ of the solitary binary with component masses $M_1$ and $M_{\rm out}$.

As expected, the distributions of final separations log\,$a_{\rm f}$ of the solitary binaries in our simulations are shifted only marginally inward compared to the initial distributions.  We find the final separation distributions peak at $\langle$log\,$a_{\rm f}$\,(AU)$\rangle$~=~1.1 and $\langle$log\,$a_{\rm f}\rangle$~=~1.6 for the B1 (initial $\mu_{\rm loga}$~=~1.5) and B2 (initial $\mu_{\rm loga}$~=~2.0) models, respectively. However, the eccentricities of the remaining solitary binaries are weighted toward very large values.  Disruptions of unstable coplanar triples produce solitary binaries that follow an eccentricity distribution of (Eqn.~7.17 in \citealt{Valtonen2006}):

\begin{equation}
 p = e(1-e^2)^{-\nicefrac{1}{2}}
\end{equation}

\noindent According to this analytic result, $F_{e>0.975}$~=~21\% of the remaining binaries have $e$~$>$~0.975, and $\approx$9\% have $e$~$>$~0.995.  Full numerical simulations conducted by \citet{Saslaw1974} confirm that $F_{e>0.975}$ = 19\%\,$\pm$\,2\% of the solitary binaries in initially unstable coplanar triples have $e^2$~$>$~0.95, i.e. $e$~$>$~0.975 (see Fig.~7.6 in \citealt{Valtonen2006}).  While the fraction $F_{e>0.975}$~$\approx$~20\% of solitary binaries with $e$~$>$~0.975 appears robust, the distribution of eccentricities above $e$~$>$~0.975 has not yet been adequately tested.   Moreover, dynamical disruption occurs on rapid timescales when there is still a massive natal disk and surrounding gaseous envelope (see below). The presence of a dense inner disk may prevent the solitary binaries from achieving such large eccentricities and small periastron separations.  We consider two scenarios.  In the optimistic scenario, we adopt the analytic results according to Eqn.~20 such that $F_{e>0.975}$~=~21\% of the solitary binaries have $e$~$>$~0.975 and the distribution above $e$~$>$~0.975 is skewed toward $e$~$\rightarrow$~1.  In the conservative scenario, we adopt Eqn.~20 for $e$~$<$~0.975, as motivated by the numerical simulations, but assume the $F_{e>0.975}$~=~21\% of highly eccentric binaries with $e$~$>$~0.975 are uniformly distributed across the interval $e$~=~0.9750\,-\,0.9999. 

In Fig.~8, we display the periastron separations $r_{\rm peri}$~=~$a$(1$-e$) of the remaining solitary binaries in both B1 and B2 models and for both optimistic and conservative scenarios. The closer initial separation distribution in our B1 model yields more solitary binaries with smaller periastron separations compared to our B2 model.  In the conservative scenario in which the final eccentricity distribution flattens above $e$~$>$~0.975, the frequency of solitary binaries with small periastron separations is reduced, but only slightly, i.e., less than a factor of two.

\begin{figure}[t!]
\centerline{
\includegraphics[trim=0.8cm 0.1cm 1.4cm 0.1cm, clip=true, width=3.7in]{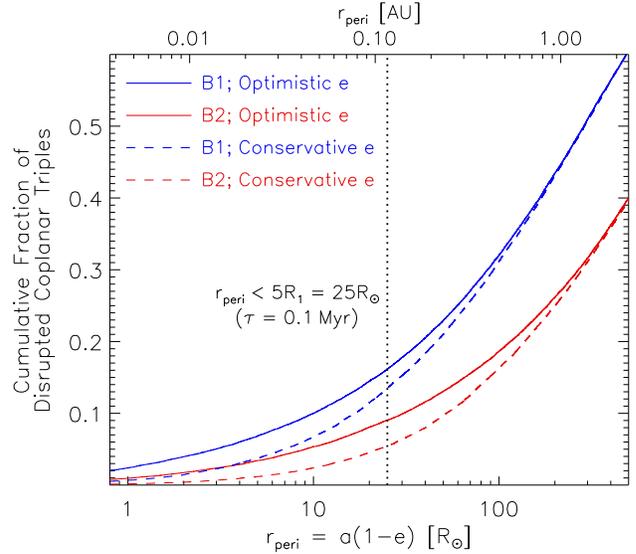}}
\caption{After disruption of unstable coplanar triples that initially fragment within the disk, we show the cumulative distribution of periastron separations $r_{\rm peri}$~=~$a$(1$-e$) of the remaining solitary binaries.  We compare the results based on different initial conditions B1 (blue; $a_{\rm in,0}$~$\sim$~$a_{\rm out,0}$~$\sim$~30~AU) and B2 (red; $a_{\rm in,0}$~$\sim$~$a_{\rm out,0}$~$\sim$~100~AU) as well as optimistic (solid) and conservative (dotted) assumptions regarding the final eccentricity distributions. Dynamical ejections occur on timescales of 10$^3$ orbits ($\sim$\,0.1\,Myr) during the early pre-MS phase when $R_1$~$\approx$~5$R_1$.  After dynamical ejections, the remaining solitary binaries that achieve $r_{\rm peri}$~$<$~5$R_1$~$\approx$~25\Rsun\ (left of dotted line; 6\%\,-\,17\% of dynamically disrupted coplanar triples) will subsequently tidally decay toward shorter periods $P$~$\approx$~10\,-\,100 days. Although this channel cannot explain very close binaries with $P$~$<$~10~days, which likely migrated at slightly later ages $\tau$~=~0.3\,-\,5~Myr when the pre-MS primary was more compact, this dynamical ejection scenario may produce a large fraction of the observed solitary binary population with $P$~$\approx$~10\,-\,100~days.}
\end{figure}

For our unstable coplanar triples with $P_{\rm in,0}$~$\sim$~$P_{\rm out,0}$~$\sim$~100\,yrs, the timescale for dynamical ejection of one of the components is $\sim$10$^3$ orbits, i.e., $\sim$\,0.1 Myr.  At such young ages, the pre-MS primary is still quite large ($R_1$~$\approx$~5\Rsun).  In Fig.~8, we indicate our criterion $r_{\rm peri}$~$<$~5$R_1$~$\approx$~25\Rsun\ for tidal energy dissipation via dynamical oscillations to sufficiently decay the remaining binary toward shorter periods $P$~$<$~100~days.  Depending on the model, we find 6\%\,-\,17\% of dynamically disrupted coplanar triples leave behind solitary binaries that achieve such small periastron separations. Assuming $F_{\rm TripleDisk}$~=~10\% of systems are initially born in coplanar, unstable triples in which both companions fragment in the disk, then 0.6\%\,-\,1.7\% of systems are solitary binaries with $P$~$<$~100~days that derive from this dynamical ejection scenario.  

Only a small fraction of the simulated systems that reach $r_{\rm peri}$~$<$~25\Rsun\ will decay to very short periods $P$~$<$~10~days.  As indicated in \S3, inner binaries that achieve small periastron separations $r_{\rm peri}$~$\approx$~5$R_1$ while the pre-MS primary is still very large, i.e., $R_1$~$\gtrsim$~5~\Rsun, will tidally decay to periods $P$~$\approx$~10\,-\,100.  The majority of close binaries with $P$~$<$~10~days must have migrated at slightly older ages $\tau$~$\approx$~0.3\,-\,5~Myr during the later pre-MS phase when there was still a disk but the primary was more compact ($R_1$~=~1\,-\,3\,\Rsun).  Most importantly, a significant majority of binaries with $P$~$<$~10~days have gravitationally bound outer tertiaries, and so could not have derived from a triple-star disruption process.  

Nevertheless, the disruption of coplanar triples in the disk may measurably contribute to the population of slightly wider solitary  binaries with $P$~=~10\,-\,100~days. About 4\% of solar-type MS stars have binary companions across $P$~=~10\,-\,100 days \citep{Raghavan2010,Moe2017}, $\approx$70\% of which do not have tertiary components \citep{Tokovinin2006}.   This implies that 4\%\,$\times$\,70\%~$\approx$~3\% of solar-type MS stars are in solitary binaries with $P$~=~10\,-\,100 days.  In our models, we find that 0.6\%\,-\,1.7\% of systems have solitary binaries with $P$~=~10\,-\,100 days that derive from the disruption of unstable triples, which is a sizable fraction of the observed value of 3\%.  Solar-type binaries with $P$~=~10\,-\,100~days also exhibit an excess twin fraction, although not as large as observed in very close binaries with $P$~$<$~10~days \citep{Tokovinin2000,Halbwachs2003,Moe2017}.  In addition, the binary fraction of pre-MS T~Tauri stars across $P$~=~10\,-\,100 days is consistent with that observed in the field MS population \citep{Mathieu1994,Melo2003,Moe2017}. These observations demonstrate that slightly wider binaries with $P$~=~10\,-\,100~days must also have originally migrated during the pre-MS phase.  Our dynamical ejection scenario, which occurs on timescales of $\sim$\,0.1~Myr, is consistent with these time constraints. Finally, solitary binaries across $P$~=~10\,-\,100~days have systematically smaller eccentricities than their counterparts with tertiary companions \citep{Raghavan2010}.  As discussed in \S3, binaries that are tidally captured into shorter orbits when $R_1$~$>$~5\,\Rsun\ will not only remain at slightly longer periods $P$~=~10\,-\,100~days, but also tidally decay toward smaller eccentricities (see Fig.~6).  We therefore expect the solitary binaries that achieve $r_{\rm peri}$~$\lesssim$~25\,\Rsun\ during the early pre-MS phase will predominantly have intermediate periods $P$~=~10\,-\,100~days and small to modest eccentricities by the zero-age MS.  Although the disruption of unstable coplanar triples in the disk cannot significantly contribute to the very close binary population with $P_{\rm close}$~$\equiv$~1\,-\,10~days, this channel may still be responsible for a non-negligible fraction of the observed population of solitary binaries with $P$~$\approx$~10\,-\,100 days. 

\section{Discussion}

\subsection{Summary of Observational Constraints}

 Nearly all close binaries have outer tertiary companions \citep{Tokovinin2006}, but this observation does not necessarily dictate close binaries migrated via KL oscillations in misaligned triples.  As emphasized in \citet{Bate2002} and \citet{Bate2009}, while close binaries may require dynamical interactions with tertiary components in order to harden toward $P_{\rm in}$~$<$~10~days, most also require significant energy dissipation in the primordial disk. At least five independent lines of observational evidence show the majority of close binaries migrated to $P_{\rm in}$~$<$~10~days during the pre-MS phase.  First, the component masses of close binaries with $P_{\rm in}$~$<$~10~days are highly correlated \citep{Tokovinin2000, Halbwachs2003, Moe2017}, indicating they co-evolved through shared accretion in the disk and/or Roche-lobe overflow during the pre-MS phase \citep{Kroupa1995,Bate1997,Tokovinin2000,Bate2000,Bate2002}.  Second, a population of circularized binaries extends beyond the MS circularization period \citep{Halbwachs2003,Meibom2005,Kirk2016}, which can only occur via tidal friction involving substantially larger pre-MS components (\S3).  Third, only a small fraction of close MS binaries with $P$~$<$~10~days have non-zero eccentricities \citep{Halbwachs2003,Meibom2005,Raghavan2010}, and so KL cycles with outer tertiaries cannot continuously populate short-period eccentric binaries during the MS.  Fourth, an indirect indication of (early) KL migration of close binaries is the absence of circumbinary planets, which would likely disrupt the planet formation process \citep{Kratter2017}, although late-stage KL migration might destroy such planetary systems as well \citep{Munoz2015a,Martin2015,Hamers2016}.  Finally (and most stringent),  the close binary fraction of pre-MS T~Tauri stars is consistent with the close binary fraction of MS stars in the field \citep{Mathieu1994,Melo2003,Moe2017}.

The measured inclination distributions of triples also demonstrate the majority of close binaries cannot derive from KL cycles. Inner binaries with $P_{\rm in}$~$<$~10~days have tertiary companions across a broad parameter space of orbital periods log\,$P_{\rm out}$\,(days)~=~1.5\,-\,8.7 (Fig.~3 in \citealt{Tokovinin2008}).  The observations are relatively complete across this parameter space, and so the logarithmic periods of the outer tertiaries are distributed relatively uniformly across the interval log\,$P_{\rm out}$\,(days)~$\approx$~2\,-\,8 with small tails on either side.  This implies that $\approx$\,50\% of close binaries with outer tertiaries are in compact configurations with log\,$P_{\rm out}$\,(days)~$<$~5 ($a_{\rm out}$~$<$~50~AU). Such compact triples derive from systems in which both companions fragment within the disk, and are therefore likely to be coplanar. Indeed, compact triples with $a_{\rm out}$~$<$~50~AU all exhibit prograde configurations, a significant majority of which have small mutual inclinations $i_{\rm tot}$~$<$~30$^{\circ}$  \citep{Sterzik2002,Borkovits2016,Tokovinin2017}. In particular, \citet{Borkovits2016} measured the mutual inclinations of compact triples with $P_{\rm in}$~$<$~10~days and $P_{\rm out}$~$<$~10$^4$~days ($a_{\rm out}$~$<$~10~AU). They found the inclination distribution skewed heavily toward coplanar configurations, where $\approx$30\% had $i_{\rm tot}$~$<$~5$^{\circ}$, $\approx$70\% had $i_{\rm tot}$~$<$~30$^{\circ}$, and there were no systems with $i_{\rm tot}$~$>$~60$^{\circ}$.  \citet{Borkovits2016} noticed a small secondary peak in the inclination distribution at $i_{\rm tot}$~$\approx$~40$^{\circ}$, which they attributed to systems that formed via KL cycles and tidal friction.  However, this enhancement near $i_{\rm tot}$~$\approx$~40$^{\circ}$ accounts for only $\approx$10\% of the systems.  Stellar-mass triples with nearly coplanar configurations do not undergo significant KL oscillations (\S2.2).   The observations demonstrate that $\approx$90\% of compact triples with $P_{\rm in}$~$<$~10~days and $P_{\rm out}$~$<$~10$^4$~days ($a_{\rm out}$~$<$~10~AU) are in nearly coplanar configurations and therefore did not derive from KL migration.

\subsection{Summary of Population Synthesis Models}

Our simulations of KL migration in triples (\S2\,-\,\S3) differ from previous population synthesis models \citep{Kiseleva1998,Fabrycky2007,Naoz2014} in two important ways.  First, the cited studies initiated their models with already close inner binaries and assumed all triples are randomly oriented. We instead assume inner binaries form via disk fragmentation at intermediate separations and that only very wide outer tertiaries that form via core fragmentation have random orientations. 

Second, previous studies incorporated tidal energy dissipation as parameterized by the weak-friction equilibrium tide model \citep{Zahn1977,Hut1981,Eggleton1998}.  During KL migration, however, inner binaries must first reach very large eccentricities and small periastron separations for tidal friction to be important, and so weak-friction equilibrium tides is no longer applicable (\S2.3). We instead implement prescriptions for tidal energy dissipation via non-radial dynamical oscillations \citep{Press1977,McMillan1986,Kochanek1992, Mardling1995, Lai1997}, which is orders of magnitude more efficient in decaying the inner binaries toward smaller separations (Fig.~4). Full numerical simulations of dynamical tides demonstrate that binaries evolve from nearly parabolic configurations toward smaller eccentricities while nearly preserving the same periastron separations $r_{\rm peri}$~$\approx$~(3\,-\,5)$R_1$ on timescales of a few hundred to a few thousand orbits \citep{McMillan1986,Kochanek1992,Mardling1995}.  Our models that include dynamical tides reproduce (1) the observed deficit of highly eccentric binaries with $e$~$>$~0.8 across intermediate periods $P$~=~10\,-\,100~days, and (2) the observed concentration of triples with moderately eccentric inner binaries extending from $e_{\rm in}$~=~0.2 at $P_{\rm in}$~=~10~days to $e_{\rm in}$~=~0.8 at $P_{\rm in}$~=~100~days.

By incorporating more realistic initial conditions, dynamical tides, and larger pre-MS primaries in our models, we find that half of close binaries which migrated to $P_{\rm in}$~$<$~10~days via KL cycles and tidal friction did so during the pre-MS phase (\S3). Dynamical tides dramatically accelerate orbital decay and widen the parameter space of inner binaries that undergo KL migration without being tidally disrupted. However, our baseline model predicts that only $\approx$40\% of the observed close binary fraction $F_{\rm close}$ = 2.1\% can derive from KL cycles and tidal friction.  Considering different model parameters (e.g., less dissipative dynamical tides and smaller fraction of systems in misaligned triples) as well as caveats not encompassed in our models (e.g., full integration of dynamical oscillations, tidal precession, and suppression of secular evolution at young ages due to enveloping gas) tend to reduce the expected fraction of close binaries that derive from KL migration.  There are simply not enough triples in the necessary configurations to explain the observed close binary fraction exclusively through secular evolution.  Instead, we expect four different channels of triple-star dynamical evolution to contribute to the formation of close binaries with $P$~$<$~10~days.  We summarize their predicted rates and properties as follows:

\begin{enumerate}

\item Only $\approx$15\% of close inner binaries with $P_{\rm in}$~$<$~10~days derive from KL oscillations and tidal friction during the pre-MS phase (Fig.~5).  These systems have outer tertiaries with mutually inclined orbits, mostly across intermediate periods log\,$P_{\rm out}$\,(days)~$\approx$~4\,-\,7 ($a_{\rm out}$~$\approx$~10\,-\,1,000~AU).   Only a small fraction of misaligned outer tertiaries extend toward shorter separations $a_{\rm out}$~$<$~10~AU, as observed in the \citet{Borkovits2016} sample. We predict that T~Tauri stars with close companions will exhibit a deficit of extremely wide tertiary companions beyond log\,$P_{\rm out}$\,(days)~$\gtrsim$~7 ($a_{\rm out}$~$\gtrsim$~1,000 AU), as triples with $a_{\rm out}$~$\gtrsim$~1,000~AU have KL timescales $\tau_{\rm KL}$~$\gtrsim$~5~Myr longer than the pre-MS phase (Fig.~5).   

\item An additional $\approx$15\% of close inner binaries with $P_{\rm in}$~$<$~10~days derive from KL oscillations and tidal friction during the MS phase (Fig.~5).  About $\approx$40\% of these systems initially decayed toward shorter periods $P_{\rm in}$~$<$~100~days via dynamical tides during the pre-MS phase, but subsequent weak-friction equilibrium tides did not harden the inner binaries to very short periods $P_{\rm in}$~$<$~10~days until after the zero-age MS (Fig.~6).  Triples that form close binaries during the MS phase have inclined wide outer tertiaries with log\,$P_{\rm out}$\,(days)~$\approx$~6\,-\,8 ($a_{\rm out}$ $\approx$ 200\,-\,5,000 AU). A slight $\approx$15\% increase in the close binary fraction between $\tau$~=~5~Myr and $\tau$~=~5~Gyr during the MS phase is consistent with the observational constraints.

\item Possibly $\approx$10\% of close binaries with $P$~$<$~10~days derive from the dynamical disruption of unstable coplanar triples that initially fragmented within the disk (\S4).  Nearly all close binaries have gravitationally bound tertiary companions, and so this channel cannot significantly contribute to the very close binary population.  However, dynamical disruption of coplanar triples may produce a large fraction of the observed population of isolated binaries with slightly longer periods $P$~=~10\,-\,100~days and small eccentricities.

\item The remaining $\approx$60\% of close binaries with $P_{\rm in}$~$<$~10~days derive from the dynamical unfolding of initially unstable triples that fragment in the disk coupled with significant energy dissipation within the disk. Close binaries formed through this channel have nearly coplanar tertiary components in compact configurations with log\,$P_{\rm out}$\,(days)~$\approx$~2\,-\,5 ($a_{\rm out}$ $\approx$ 0.5\,-\,50 AU). This channel produces close binaries exclusively during the pre-MS phase while there is still dissipative gas in the primordial disk. Interactions of coplanar triples embedded in disks, not secular evolution of misaligned triples and not disk migration of solitary binaries, are required to explain the majority of very close binaries.

\end{enumerate}

\subsection{Applications to Hot Jupiters}

Tidal evolution involving planetary companions is even less well understood.  The degree of tidal energy dissipation via dynamical oscillations within the primary star scales as $\Delta E_{\rm tide}$~$\propto$~$M_2^2 (r_{\rm peri,in}/R_1)^{-9}$ (see Eqn.~9).  For stellar-mass companions with $M_2$~$=$~0.1\,-\,1.0\Msun, we showed in \S2 that inner binaries that achieve small periastron separations $r_{\rm peri,in}$~$\lesssim$~5$R_1$ effectively decay toward shorter separations via dynamical tides.  To match the same levels of tidal energy dissipation within the primary stars, however, Jupiter-mass planets with $M_2$~=~$M_J$~$\approx$~10$^{-3}$\,\Msun\ would have to reach extremely small periastron separations $r_{\rm peri,in}$~$<$~1.6$R_1$. This is inside the Roche limit of the primary stars, and so Jupiter-mass planets that achieve such small periastron separations due to KL oscillations would be tidally disrupted. Various population synthesis studies have not only simulated the formation of hot Jupiters, but also predict a non-negligible fraction of Jupiters will reach small enough periastron separations to become tidally disrupted \citep{Naoz2012,Munoz2016}.  

Dynamical oscillations within the interiors of the primary stars may therefore not play a significant role in the formation of hot Jupiters.  Most dynamical formation models of hot Jupiters find that tidal evolution is instead dominated by energy dissipation within the more compact Jupiter-mass planets ($R_J$~$\approx$~0.1\Rsun), {\it not} the stellar interiors \citep{Fabrycky2007,Naoz2012,Munoz2016}. For weak-friction equilibrium tides involving solar-mass $R_1$~=~1\Rsun\ MS primaries, tidal energy dissipation within the Jupiter-mass planets, which scales as $\Delta E_{\rm tide}$~$\propto$~$M_1^2 R_J^5$, is indeed larger than tidal energy dissipation with the primary stars, which scales as $\Delta E_{\rm tide}$~$\propto$~$M_J^2 R_1^5$.  For pre-MS primaries with larger radii $R_1$~$\approx$~2\,-\,10\Rsun\ and larger apsidal motion constants $k_1$~$\approx$~0.13 \citep{Claret2012}, however, tidal energy dissipation within the stellar interiors is orders of magnitude more efficient.  During the pre-MS phase,  energy dissipation via weak-friction equilibrium tides may actually be greater in the primary stars than in the planets.  Finally, while dynamical oscillations within the stars may not contribute to the tidal evolution of hot Jupiters, dynamical oscillations within the planetary interiors may dramatically assist in the high-eccentricity  KL migration of hot Jupiters  \citep{Ivanov2004, Chernov2017,Wu2017, Vick2017}.  To reliably predict the formation rates and properties of hot Jupiters, we must first better understand tidal energy dissipation via both dynamical oscillations and weak-friction equilibrium tides within MS stars, pre-MS stars, and planetary interiors.

M.M. acknowledges financial support from NASA's Einstein Postdoctoral Fellowship program PF5-160139. We thank the anonymous referee for insightful comments and suggestions, and we thank D.~Mu{\~n}oz and K.~Anderson for enlightening discussions and helpful feedback.

\section*{Appendix: Equations for Precession} 

A dimensionless parameter that describes the relative strength of tidal precession relative to the quadrupole level is given by (see Eqn.~35 in \citealt{Liu2015}):

\begin{equation}
 \epsilon_{\rm tide} = \frac{15 M_1 (M_1 + M_2)a_{\rm out}^3(1-e_{\rm out}^2)^{\nicefrac{3}{2}}k_{\rm 1, Love} R_1^5}{a_{\rm in}^8 M_2 M_3}
\end{equation}

\noindent where $k_{\rm 1, Love}$ is the tidal Love number of the primary. Similarly, the relative strength of precession due to rotation-induced oblateness is (see Eqn. 39 in \citealt{Liu2015}):

\begin{equation}
 \epsilon_{\rm rot} = \frac{(M_1 + M_2)a_{\rm out}^3(1-e_{\rm out}^2)^{\nicefrac{3}{2}} k_1 \Omega_1^2 R_1^5}{G a_{\rm in}^5 M_2 M_3}
\end{equation}

The characteristic KL rate is:

\begin{equation}
 \dot{\omega}_{\rm KL} = \frac{1}{\tau_{\rm KL} \sqrt{1 - e_{\rm in}^2}},
\end{equation}

\noindent  the tidal precession rate is:

\begin{equation}
 \dot{\omega}_{\rm tide} = \frac{\epsilon_{\rm tide}}{\tau_{\rm KL}}\frac{1+3e_{\rm in}^2/2 + e_{\rm in}^4/8}{(1-e_{\rm in}^2)^5}, 
\end{equation}

\noindent and the rotational precession rate is:

\begin{equation}
\dot{\omega}_{\rm rot} = \frac{\epsilon_{\rm rot}}{\tau_{\rm KL}}\frac{1}{(1-e_{\rm in}^2)^2}
\end{equation}

\noindent (see Eqns. 55, 43, and 44, respectively, in \citealt{Liu2015}).  \citet{Liu2015} find that the eccentricity is limited by tidal precession when $\dot{\omega}_{\rm tide}$ $\approx$ 7$\dot{\omega}_{\rm KL}$ and is limited by rotational precession when $\dot{\omega}_{\rm rot}$ $\approx$ 3$\dot{\omega}_{\rm KL}$.  For any particular system, we can use these analytic relations to solve for the maximum eccentricity $e_{\rm in,lim}$ limited by precession and the corresponding minimum periastron separation $r_{\rm peri,lim}$.

For instance, the inner binary in our example triple shown in Fig.~4 reached a minimum periastron separation of $r_{\rm peri,in}$ = 2.9$R_1$ due to dynamical tides during its third KL cycle when $\tau$~$\approx$~0.6~Myr and $R_1$~=~2.2\,\Rsun.  At this age, the apsidal motion constant of the primary is $k_1$~$\approx$~0.1, and so the tidal Love number is $k_{\rm 1, Love}$~=~2$k_1$~=~0.2.  The average rotational period of 1\,\Msun\ pre-MS stars is $P_{\rm rot}$~$\approx$~3~days, but the fastest 10\% rotate at $P_{\rm rot}$ $\approx$ 0.6 days \citep{Mellon2017}.  To maximize the effect of rotational precession, we assume $P_{\rm rot}$ = 0.6 days, i.e., $\Omega_1$~=~1.2$\times$10$^{-4}$~rad~s$^{-1}$. Solving for $\dot{\omega}_{\rm tide}$ $\approx$ 7$\dot{\omega}_{\rm KL}$, the limiting eccentricity due to tidal precession is $e_{\rm in,lim}$~=~0.9994, corresponding to $r_{\rm peri,lim}$~=~1.6$R_1$.  At this eccentricity, the rotational precession rate is $\dot{\omega}_{\rm rot}$~$\approx$~1.3$\dot{\omega}_{\rm KL}$.  Even for a rapidly rotating pre-MS primary with $P_{\rm rot}$~=~0.6~days, tidal precession limits the eccentricity before rotation-induced oblateness becomes important.  The minimum periastron separation $r_{\rm peri,lim}$~=~1.6$R_1$ limited by tidal precession is well inside the tidal disruption threshold $r_{\rm peri,in}$~$\approx$~2.5$R_1$.  In the absence of tidal friction, the inner binary in our example triple would have been tidally disrupted long before tidal precession could have limited the eccentricity. We repeat this calculation for our baseline population of triples, and the results are presented in \S2.3.5.

\bibliographystyle{apj}                       %% AASTeX
\bibliography{bibliography}

\end{document}